\documentclass[%
 reprint,
superscriptaddress,
 amsmath,amssymb,
aps,
]{revtex4-1}

\newcommand{\iflanggerman}[2]{
 \iflanguage{german}{#1}{
  \iflanguage{ngerman}{#1}{#2}
 }
}

\iflanggerman{

}{

}

\DeclareMathOperator{\tr}{tr}

\DeclareMathOperator{\im}{Im}

\newcommand{\intinf}{\int\limits_{-\infty}^{+\infty}}
\DeclareMathOperator\erf{erf}
\DeclareMathOperator\Var{Var}
\DeclareMathOperator\E{E}
\DeclareMathOperator\Prob{P}

\usepackage{graphicx,xcolor}% Include figure files
\usepackage{dcolumn}% Align table columns on decimal point
\usepackage{bm}% bold math
\usepackage[normalem]{ulem}
\usepackage{bbm}% bold math
\usepackage{hyperref}% add hypertext capabilities
\usepackage[capitalise]{cleveref}
\usepackage{siunitx}
\usepackage{mathtools}
\usepackage[utf8]{inputenc}

\usepackage{amsfonts}
\usepackage{amstext}
\usepackage{amsmath}
\usepackage{amsthm}
\usepackage{amssymb}
\usepackage{amsbsy}   % AMS-Boldsymbol

\newcommand{\newx}[0]{\xi(x)}
\newcommand{\sigmaU}[0]{\sigma_{\bar l}\sqrt{N-2}}
\newcommand{\dlc}[0]{\Delta l_i|_{\bar l_i=\barell}}
\newcommand{\gc}[0]{g|_{\bar l_i=\barell}}
\newcommand{\barlst}[0]{\bar l_\text{st}}
\newcommand{\supr}[0]{\sup_{x \in \mathbb R}}
\newcommand{\dlC}[0]{\Delta l_i|\bar l_i=\barell}
\newcommand{\sn}[1]{{\color{black}{#1}}}

\newcommand{\lbar}[0]{{\boldsymbol{\bar l}}}
\newcommand{\barz}[0]{z}
\newcommand{\lstar}[0]{\boldsymbol{l^*}}
\newcommand{\ellstar}[0]{{\ell^*}}

\newcommand{\ellbar}[0]{{\bar \ell}}

\newcommand{\barell}[0]{{\bar \ell}}
\newcommand{\dl}[0]{\boldsymbol{\Delta l}}
\newcommand{\dell}[0]{{\Delta \ell}}
\newcommand{\dlgiven}[0]{{\dl|\lbar=\bar \ell}}
\newcommand{\vardl}[0]{{\Var(\dl)}}
\newcommand{\varlbar}[0]{{\Var(\bar l)}}
\newcommand{\vardlgiven}[0]{{\Var(\dl|\lbar)}}
\newcommand{\vardlgivenl}[0]{{\Var(\dl|\lbar=\bar\ell)}}
\newcommand{\meanvar}[0]{{\E_{\lbar}[\vardlgiven]}}
\newcommand{\A}[0]{\mathbf{A}}
\newcommand{\Q}[0]{\mathbf{Q}}
\newcommand{\Per}[0]{\mathbf{U}}
\newcommand{\C}[0]{\mathbf{C}}
\newcommand{\CT}[0]{\mathbf{\tilde C}}

\newtheorem*{thm}{Theorem}

\hypersetup{pdfauthor={Knut Heidemann}, pdftitle={Topology determines force distributions in one-dimensional random spring networks},pdfkeywords={force distribution, topology, random graphs, spring network}}
\let\emph\relax % there's no \RedeclareTextFontCommand
\DeclareTextFontCommand{\emph}{\textnormal\em}

\newcommand{\mw}[1]{\textcolor{black}{#1}}

\begin{document}

%\preprint{APS/123-QED}

%\title{Loops count: force distributions in circular spring networks}% Force line breaks with \\
%\title{Topology matters: force distributions in circular spring networks}% Force line breaks with \\
%\title{\sn{Topology counts}: force distributions in circular spring networks}% Force line breaks with \\
%\title{A graph-theoretic theory of random spring networks on the circle}% Force line breaks with \\
%\title{Ensembles of random spring networks on the circle and their force distributions}% Force line breaks with \\
\title{Topology determines force distributions in one-dimensional random spring networks}% Force line breaks with \\
%\title{Towards a topological understanding of random spring networks}% Force line breaks with \\
%\title{Topological (d)ef(f)ects on forced random spring networks}% Force line breaks with \\
%\title{A topological perspective on random spring networks}% Force line breaks with \\

\author{Knut M. Heidemann}
\thanks{These two authors contributed equally.}
%\email{k.heidemann@math.uni-goettingen.de}
\affiliation{Institute for Numerical and Applied Mathematics, University of Goettingen, Germany}%Lines break automatically or can be forced with \\
\author{Andrew O. Sageman-Furnas}
\thanks{These two authors contributed equally.}
%\email{k.heidemann@math.uni-goettingen.de}
\affiliation{Institute for Numerical and Applied Mathematics, University of Goettingen, Germany}%Lines break automatically or can be forced with \\
\author{Abhinav Sharma}%
\affiliation{Third Institute of Physics---Biophysics, University of Goettingen, Germany}%

\author{Florian Rehfeldt}
\affiliation{Third Institute of Physics---Biophysics, University of Goettingen, Germany}%

\author{Christoph F. Schmidt}
\email{cfs@physik3.gwdg.de}
\affiliation{Third Institute of Physics---Biophysics, University of Goettingen, Germany}%

\author{Max Wardetzky}
\email{wardetzky@math.uni-goettingen.de}
\affiliation{Institute for Numerical and Applied Mathematics, University of Goettingen, Germany}%Lines break automatically or can be forced with \\

\date{\today}% It is always \today, today,


\begin{abstract}
%We study the distribution of forces in random spring networks by introducing a \emph{full network} model on the circle. A network with internal forces relaxes into mechanical equilibrium with an inhomogeneous force distribution.
%We show that one needs to go \emph{beyond} traditional \emph{mean-field} approaches to characterize the resulting force distributions.
%Using \emph{graph-theoretic} techniques we derive the distributions and illustrate the essential role network topology plays in the system.
%Our analytical approach is straightforwardly generalized to higher spatial dimensions.
Networks of elastic fibers are ubiquitous in biological systems and often provide mechanical stability to cells and tissues.
Fiber reinforced materials are also common in technology.
An important characteristic of such materials is their resistance to failure under load. Rupture occurs when fibers break under excessive force and when that failure propagates. Therefore it is crucial to understand force distributions.
%govern the mechanical properties of many natural as well as man-made materials.
%Force distributions within these networks can be highly inhomogeneous and, although the importance of force distributions for structural properties is well recognized, they are far from being understood quantitatively.
Force distributions within such networks are typically highly inhomogeneous and are not well understood.
%In this study we show for the first time that full \emph{network topology} beyond the mean-field approach is essential for accurately capturing force distributions.
Here we construct a simple one-dimensional model system with periodic boundary conditions by randomly placing linear springs on a circle. We consider ensembles of such networks that
consist of $N$ nodes and have an average degree of connectivity $z$, but vary in topology.
%We study force distributions in a model system consisting of an ensemble of random linear spring networks on a circle that
%%share a common number of nodes $N$ and average degree of connectivity $z$,
%\mw{consist of $N$ nodes and have an average degree of connectivity $z$, but vary in topology.}
%Using a graph-theoretical approach that accounts for the full topology of each network in the ensemble, we show that, surprisingly, the force distributions can be characterized only in terms of the parameters $(N,z)$.
Using a graph-theoretical approach that accounts for the full topology of each network in the ensemble, we show that, surprisingly, the force distributions can be fully characterized in terms of the parameters $(N,z)$.
Despite the universal properties of such $(N,z)$-ensembles, our analysis further reveals that a classical mean-field approach fails to capture force distributions correctly.
We demonstrate that \emph{network topology} is a crucial determinant of force distributions in elastic spring networks.
\end{abstract}
\maketitle

\graphicspath{ {../} {../img/}}

%might look like a direct application. however, since the edge lengths are not indep. 

% section Assumptions etc. (end)
%\section{Introduction}
%\sn{Do we need this paragraph?}
%\sm{Add some math to intro/discussion.}
\section{Introduction} % (fold)
\label{sec:Introduction}

% section Introduction (end)
Networks---and their \emph{topologies}---have been studied in a broad range of disciplines, leading to terms like \emph{social, economic, biological}, or \emph{chemical} networks, and, of course, mechanical networks \cite{newman2010networks}.
Here we focus on the latter and expand the theoretical and numerical analysis introduced in a companion short paper \cite{heidemann2017b}. Networks of filamentous proteins, polysaccharides or nucleic acids, essentially all semiflexible filaments, play important roles for the mechanics and stability of biological cells and tissues \cite{MacKintosh1997,howard2001mechanics}.
An important design feature of biological materials is the response to large loads, including failure, rupture, damage limitation and their recovery properties.
To understand failure that starts with the rupture of single filaments when the local force exceeds a threshold, it is crucial to understand force distributions in filament networks.
It turns out that topology plays a critical role for the distribution of forces in elastic (e.g, polymer) networks, but this topic has received little attention to date.

The quantitative analysis of force distributions within random polymer networks has largely relied on computational modeling \cite{heussinger2007force,Heidemann2014}.
%Network deformation is highly non-affine on the microscopic scale\cite{onck2005alternative,heussinger2006floppy,sharma2015strain} making an analytical treatment of the deformation field difficult. 
%Even without loading, a network can be internally stressed\cite{oosten2016uncoupling} resulting in a broad distribution of forces.
Analytical descriptions of fiber networks have primarily used \emph{effective-medium} \cite{Thorpe1985,Broedersz2011b,Sheinman2012a} or \emph{mean-field} \cite{Storm2005,Sharma2013a,Heidemann2014} approaches.
Effective-medium theories rely on mapping a disordered system to an ordered one.
%, while requiring a self-consistency assumption as well as constraints on the local displacement fields.
It is unclear, however, how force distributions change under this mapping.
%In fact, effective-medium approaches rather focus on predicting the \emph{macroscopic} response, rather than properties of the \emph{microscopic} displacement fields.
%\sm{which either neglect network topology \cite{Heidemann2014} or rely on lattice-based formulations} \cite{Thorpe1985,Sheinman2012a}.
%which 
Mean-field approaches do not consider the full network topology, but only the \emph{local} degree of connectivity. %\emph{local} degree of connectivity.
%\sn{Mean-field models consider local degree of connectivity to \dots approximation to global network properties.}
We show that such an approach fails to describe force distributions even for a very simple model system; %in fact, \emph{nonlocal} dependencies introduced by cycles/loops in the network must also be considered.
%\sn{The reason for this is that topological constraints cause global coupling that remains prevalent even when the system becomes large.}
\sn{in fact, topological features, i.e., cycles/loops in the networks, cause global coupling that remains prevalent even when the system becomes large.}

\begin{figure}
  \centering
  %\includesvg[width=\columnwidth]{circle-springs-curved}
  %\includegraphics[width=\columnwidth]{circle-springs-curved}
  %\showthe\columnwidth % Use this to determine the width of the figure.
  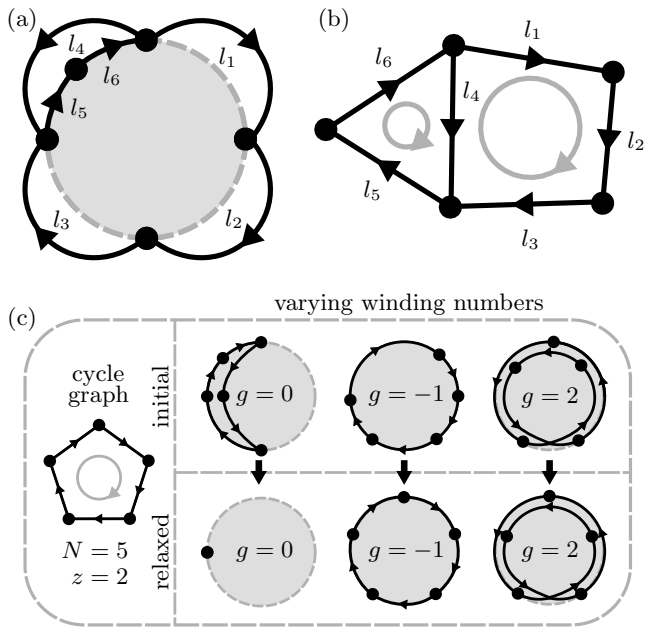
  \caption{(a) An example network on the circle, with $N=\num{5}$ and $z=\num{2.4}$. (b) Graph representation of the network in (a). The edge/spring orientations are depicted by black arrows. The network contains two fundamental cycles, for example: $\{l_1,l_2,l_3,l_4\}$ and $\{l_4,l_5,l_6\}$. After choosing arbitrary orientations for both cycles (gray arrows), we construct linear constraints that fix their winding numbers (\cref{eq:totalEnergy})---here: $l_1+l_2+l_3-l_4 = -1$ (winds around circle once) and $l_4+l_5+l_6 = 0$ (contractible). \sn{(c) The abstract cycle graph ($z=2$) with $N=5$ (left) and three realizations on the circle with distinct topologies (same graphs but different winding numbers $g$). Top and bottom row show initial and corresponding relaxed configurations, respectively. Note that, for visualization purposes, overlapping springs are drawn with a slight offset.}}
  \label{img:schematic}
\end{figure}

% section Network model (end)
The simple model system that we consider here consists of ensembles of one-dimensional random spring networks on a circle.
%ensembles of \emph{circular spring networks} \sm{as a minimal model for random fiber networks}.
Considering such networks is equivalent to applying \emph{periodic boundary conditions} in one dimension.
To model the effect of external force applied to the network, we employ a generation procedure that inserts springs with pre-strain so that the resulting initial configurations are not in mechanical equilibrium.
We then study the resulting force distributions of the relaxed systems.

%\section{Network generation} % (fold)
We generate \emph{initial} network configurations as follows (\cref{img:schematic}):
(i) Place $N$ node positions \sn{(indexed from 1 to $N$)} drawn from a \emph{uniform} distribution on the circle.
(ii) Connect these nodes in the order given by their indices into one connected \emph{cycle} via springs.
We always connect consecutive nodes via the \emph{shorter} of the two possible distances.
\sn{Note that the cycle may wrap around the circle zero, one, or multiple times (\cref{img:schematic}~(c)).}
This step guarantees that each network will always have only one connected component \sn{and prevents dangling ends}.
(iii) Connect further node pairs \emph{randomly}, such that each node pair is connected by at most one spring, until the network contains $Nz/2$ springs, where the \emph{average degree} of \sn{connectivity} $\barz$ is chosen such that $Nz/2$ is an integer.

Each spring is linear, has rest length zero, and unit spring constant. Its length is measured along the circumference of the circle. 
%We choose as degrees of freedom the \emph{signed} spring lengths in units of the circle's circumference (positive sign means nodes are connected counter-clockwise, and vice versa).
In order to encode this construction in an unambiguous manner we work with signed spring lengths as degrees of freedom.
\sn{The orientation of a spring is chosen such that it goes from a node of lower index to a node of higher index. This is an arbitrary choice, but defined orientations are essential in our formalism. The sign of the spring length is chosen to be positive if its orientation on the circle points counter-clockwise and negative otherwise.}
%\sm{Note that step (ii) guarantees that each network has only one connected component.}
%and the assumption of unit spring constant means lengths and forces are synonymous}.

%, i.e., $|l_{ij}|\leq0.5$.
\sn{
The network can be encoded within a \emph{graph representation}, where the springs \sn{together with their orientations} are the \emph{directed} edges of the graph, with signed lengths as edge weights (\cref{img:schematic}~(b)).
To lie on the circle, the graph and edge weights must be compatible in the sense that the sum of the edge weights around each cycle of the graph is equal to an integer, which we refer to as its winding number $g$.
%The edge direction indicates which of the two nodes is the starting point for the length measurement (see~\cref{img:schematic}~(b)).
Our network generation procedure guarantees this compatibility. It results in a \emph{random directed Hamiltonian graph}, i.e., 
a graph that contains a cycle that visits each node exactly once, with $N$ nodes and average degree $z$. This graph comes equipped with compatible initial spring lengths/edge weights $\{\bar l_i\}_{i=1}^{Nz/2}$ that are each \emph{uniformly} distributed as $\mathcal U(-0.5,0.5)$, but, since they are coupled by integer winding numbers, not mutually independent \cite{weiss2006course} as random variables.}
\sn{
We seek to characterize the length (i.e. force) distributions of springs in networks after they have relaxed to mechanical equilibrium. Relaxation preserves network topology, i.e., it preserves its graph together with a set of winding numbers, that arise from the generation process. Note that networks sharing the same graph may have different sets of winding numbers, and therefore distinct relaxed states (\cref{img:schematic}~(c)).
A particular realization of an initial network uniquely determines network topology and results in a known linear solution operator for the respective mechanical equilibrium.
However, a network ensemble, with a given connectivity and number of nodes includes many topologies. This leads to a random solution operator, which makes it more difficult to determine the ensemble-averaged distribution of relaxed lengths.

Motivated by experiments, where explicit information on particular realizations is hard to obtain, we study ensembles with a fixed number of nodes $N$ and average degree $z$, henceforth called $(N,z)$-ensembles.
Surprisingly, such ensembles have well defined force distributions despite varying topologies.
Explicitly accounting for these unknown underlying topologies makes our approach different from a mean-field description.
}

%\teal{While \cref{eq:kkt} allows for computing the equilibrium spring lengths ($\lstar$) for a specific realization by solving a linear system, it requires knowledge of the exact network topology ($\mathbf S$) and the initial edge lengths ($\lbar$);
%this information is usually not available in experiments, more accessible parameters are the average connectivity $z$ and the number of nodes $N$.
%In the following, we thus study ensembles of networks characterized by $N$ and $z$ but with varying network topology (underlying graphs).
%It is a priori unclear whether ensemble properties of the $S$ matrix as a random variable can be derived.
%Even in the simplest ensemble ($z=2$), namely the cycle graph, where network topology, hence $\mathbf S$, is determined, \sm{ now explain $\lbar$ dependencies}.}
%In the following, we study \emph{ensembles} of random networks by running our generation procedure multiple times with fixed number of nodes $N$ and average degree $z$. Note that these networks generally do not share the same underlying graph.
%We characterize ensemble averaged force distributions based on $N$ and $z$ only.
%The challenge lies in 

\section{Analytical theory} % (fold)
\label{sec:Analytical theory}
%\subsection{Problem formulation} % (fold)
%\label{sub:Problem formulation}
% subsection Problem formulation (end)
% section Analytical theory (end)
Formally, as already described in \mw{\cite{heidemann2017b}}, our model can be described as the following optimization problem:
%Finding the final spring length distribution can be phrased as the following optimization problem:}
\begin{align}
  \text{minimize}\quad &\frac{1}{2}\boldsymbol l^T \boldsymbol l\quad \text{subject to}\quad \mathbf C \boldsymbol l = \mathbf g = \mathbf C \lbar\,,
  \label{eq:totalEnergy}
\end{align}
%\sn{Knut: Can we shorten the paragraph between (1) and (2)?}
where $\boldsymbol l \in \mathbb R^{Nz/2}$ is the vector of all spring lengths and $\mathbf g \in \mathbb Z^m$ is the vector of winding numbers, which is determined by the vector of initial spring lengths $\lbar$ and the \emph{signed cycle matrix} $\mathbf C \in \mathbb Z^{m \times Nz/2}$, described below.

The first part in \cref{eq:totalEnergy} minimizes the total elastic energy of the system, whereas the second part preserves the topology of the network by fixing the winding numbers of a set of $m=N(z/2-1)+1$ \emph{fundamental cycles}.
A fundamental cycle is defined as a cycle that occurs when adding a single edge to a spanning tree of the graph. There are $N-1$ edges in the spanning tree, so $Nz/2 - (N-1)$ edges can be added. Therefore, there are $N(z/2-1)+1$ fundamental cycles.
Note that the choice of fundamental cycles corresponds to the choice of a basis and is therefore not unique.
The solution to \cref{eq:totalEnergy}, however, is independent of this choice (\cref{sec:Independence of the solution on the choice of the cycle basis}).
%imposing additional \emph{contractibility constraints}.
%It is determined by the vector of initial spring lengths $\lbar$ and the \emph{signed cycle matrix} $\mathbf C \in \mathbb Z^{m \times Nz/2}$
%, then every network would collapse to $N$ points lying on top of each other with all spring lengths 0.
%Instead, we want to preserve the circular structure of the embedding by imposing additional \emph{contractibility constraints}. %\sm{that ensure preservation of network topology (no spring rupture) throughout the minimization process}.
%However, the network being embedded onto the circle imposes additional \emph{boundary conditions}.
%Network connectivity can be readily encoded within a \emph{graph representation} --- the springs being the edges of the graph.
%In order to fix the cycle contractibility of the network we first determine a set of $m=N(z/2-1)+1$ \emph{fundamental cycles} 
%\footnote{A fundamental cycle is defined as a cycle that occurs when adding a single edge to a spanning tree of the graph. There are $N-1$ edges in the spanning tree, so $Nz/2 - (N-1)$ edges can be added. Hence, there are $N(z/2-1)+1$ fundamental cycles.} of the graph.
%\sn{Each cycle in the graph may either be contractible or not---the latter meaning that it winds around the circle at least once (see~\cref{img:schematic}).}

After choosing a cycle basis, the $\mathbf C$-matrix is constructed by specifying an orientation for each fundamental cycle and then setting $C_{ji}$ equal to: $1$ if spring $i$ is part of the $j$th fundamental cycle and their orientations agree, or $-1$ if their orientations are opposite, and $0$ otherwise.
For the example in \cref{img:schematic}~(a), the cycle matrix and vector of winding numbers are given by $C_{1}=(1,1,1,-1,0,0)$, $C_{2}=(0,0,0,1,1,1)$, and $\boldsymbol g=(-1,0)^T$, respectively.
%(see \cref{img:schematic} for an example).
Note that winding numbers correspond to the signed number of times a cycle wraps around the circle. Contractible cycles have winding number zero.
If all cycles were contractible, then \cref{eq:totalEnergy} would have a trivial solution with all springs collapsed to a single point. It is only the presence of nontrivial cycle constraints that prevents this outcome.

It is noteworthy to point out that the problem presented above is \emph{equivalent} to the classical problem of determining the currents (here $\boldsymbol l$) in an \emph{electrical network}. Force balance (or minimizing the energy $\boldsymbol l^T \boldsymbol l$) is equivalent to Kirchhoff's current law (signed currents add up to zero at a node) and---assuming unit resistances---the cycle constraints ($\mathbf C \boldsymbol l = \boldsymbol g$) correspond to Kirchhoff's voltage law (voltages in a closed loop sum up to zero), where the winding numbers $g_j$ represent voltage sources.
\Cref{eq:totalEnergy} defines a \emph{quadratic programming problem} with a unique analytic solution:
%written in terms of the spring length changes $\dl$ during relaxation to the final configuration $\lstar$, it reads:
%determined via a linear system:
%\begin{align}
%  \begin{pmatrix}
%    \boldsymbol I & -\mathbf C^T \\
%    \mathbf C & \mathbf 0
%  \end{pmatrix}
%  \begin{pmatrix}
%    \boldsymbol l^* \\ \boldsymbol \lambda^*
%  \end{pmatrix}
%  =
%  \begin{pmatrix}
%    \mathbf 0 \\ \mathbf g
%  \end{pmatrix}\,,
%  \label{eq:kkt}
%\end{align}
%with $\mathbf l^*$ and $\boldsymbol \lambda^*$ being the solution vector of spring lengths and a set of Lagrange multipliers, respectively.
\begin{align}
  \lstar &=\mathbf C^T(\mathbf C\mathbf C^T)^{-1}\underbrace{\mathbf C  \lbar}_{= \boldsymbol g} \eqqcolon \mathbf P \lbar\,,
  %\Leftrightarrow \quad \dl &\coloneqq \lstar - \lbar = \underbrace{(\mathbf P - \mathbf I)}_{\eqqcolon \mathbf S}\, \lbar\,,\\
  %\dl \coloneqq \lstar - \lbar &= (\underbrace{\mathbf C^T(\mathbf C\mathbf C^T)^{-1}\mathbf C - \mathbf I)}_{\mathbf \eqqcolon \mathbf S} \lbar\,,
  \label{eq:kkt}
\end{align}
%The problem \cref{eq:kkt} is uniquely determined by the choice of $\mathbf C$ and $\mathbf g$. 
which can be explicitly computed for each realization via, e.g., the optimization library IPOPT \cite{Wachter2006}.

To express the resulting force distributions of an $(N,z)$-ensemble we consider the expected histogram of the vector $\lstar$ of random variables.
This results in a \emph{univariate} probability density for the final spring lengths.
For a particular realization,
%(could be the initial $\lbar$ or final $\lstar$ or arbitrary),
the corresponding \emph{cumulative histogram} $H_{\lstar}$ is given via
\begin{align}
  H_{\lstar}(\ellstar) \coloneqq \frac{2}{Nz} \sum_{i=1}^{Nz/2}\mathbbm 1_{l_i^* \leq \ellstar}\,,
\end{align}
where $\mathbbm 1_{A}$ is the indicator function (one if $A$ is true, zero otherwise).
The quantity $H_{\lstar}(\ellstar)$ measures the number of elements in $\lstar$ with values less than or equal to $\ellstar$.
We are interested in a univariate \emph{cumulative distribution function} (cdf) $F_{\lstar}$, which we define as the expected value of the cumulative histogram of an $(N,z)$-ensemble:
\begin{equation}
  \begin{split}
  F_{\lstar}(\ellstar) \coloneqq& \E [H_{\lstar}(\ellstar)] = \frac{2}{Nz} \sum_{i=1}^{Nz/2} \E [\mathbbm 1_{l_i^*\leq \ellstar} ]\\
  =& \frac{2}{Nz} \sum_{i=1}^{Nz/2} \Prob(l_i^* \leq \ellstar) = \frac{2}{Nz} \sum_{i=1}^{Nz/2} F_{l_i^*} (\ellstar)\,.
  \label{eq:cumcdf}
\end{split}
\end{equation}
The quantity $F_{\lstar}(\ellstar)$ is the average over the marginal distribution functions of the individual $l_i^*$. This result defines the corresponding univariate \emph{probability density} (expected histogram), i.e.,
\begin{align}
  p_{\lstar}(\ellstar) \coloneqq \frac{d}{d\ellstar}F_{\lstar}(\ellstar)= \frac{2}{Nz} \sum_{i=1}^{Nz/2} p_{l_i^*} (\ellstar)\,.
  \label{eq:averagedDensity}
\end{align}
By decomposing the final length vector $\lstar$ into initial lengths $\lbar$ and length changes $\dl$, i.e., $\lstar = \lbar + \dl$, we compute
\begin{align*}
  p_{l_i^*}(\ell^*) = p_{\bar l_i + \Delta l_i}(\ell^*) = \intinf p_{\bar l_i }(\bar \ell) \cdot p_{\Delta l_i|\bar l_i = \bar \ell}(\ell^* - \bar \ell)\, d\ellbar\,,
\end{align*}
and therefore with \cref{eq:averagedDensity}:
\begin{align}
  \label{eq:plfgeneral}
  p_{\lstar}(\ellstar) %&= \frac{2}{Nz} \sum_{i=1}^{Nz/2} p_{l_i^*} (\ellstar)\nonumber \\
  &= \frac{2}{Nz}\sum_{i=1}^{Nz/2}\intinf p_{\bar l_i }(\bar \ell) \cdot p_{\Delta l_i|\bar l_i = \bar \ell}(\ell^* - \bar \ell)\, d\ellbar\,.
\end{align}
%for the final spring lengths we look at the displacements from the initial configuration, i.e., write 
%Moreover, we define the changes in spring lengths $\dl$ after relaxation into mechanical equilibrium via ${\lstar = \lbar + \boldsymbol{\Delta l}}$.
%It follows that
%where $\boldsymbol{\Delta l}$ are .
%an \emph{individual spring} with index $i$ is given as
%\begin{align}
  %\begin{split}
  %p_{\lstar}(\ellstar) = p_{\lbar + \dl}(\ellstar)% = \intinf p_{\lbar, \dl} (\ellbar,\ellstar-\ellbar) \,d\ellbar\\
  Remember that the initial spring lengths $\bar l_i$ are identically distributed, i.e., $p_{\bar l_i} = p_{\bar l}$.
  \Cref{eq:plfgeneral} thus simplifies to:
  \begin{align}
  \label{eq:plf}
    p_{\lstar}(\ellstar)= \intinf p_{\bar l}\, (\ellbar) \cdot p_{\dl|\lbar=\ellbar}\, (\ellstar-\ellbar) \,d\ellbar\,,\\
  \text{with}\quad p_{\dl | \lbar=\bar \ell}\,(\dell) \coloneqq\frac{2}{Nz} \sum_{i=1}^{Nz/2} p_{\Delta l_i | \bar l_i=\bar \ell}\,(\dell)\,.
  \label{eq:averagep}
\end{align}
%\end{split}
%where $p_{\dl | \lbar}(x) \coloneqq \frac{2}{Nz} \sum_{i=1}^{Nz/2} p_{\Delta l_i | \bar l_i}(x)$.
%leading to the density $p_{\lstar}$ for the full final spring vector via \cref{eq:averagedDensity}.
%We saw that $\bar l_i \sim \mathcal U(-0.5,0.5)$ and therefore $\lbar \sim \mathcal U(-0.5,0.5)$.
%\so{Note that the conditional probability debsity \cref{eq:averagep} describes the expected conditional histogram of $\dl$ given that each $\bar l_i$ is conditioned on the same value $\bar \ell$.}
Note that $p_{\dl | \lbar=\bar \ell}$, with the apparent dimensionality mismatch, is a shorthand notation that does not mean that $\bar l_i=\bar \ell$ for all indices $i$, but instead, corresponds to the average over all possible events that $\bar l_i=\bar \ell$ for some index $i$. In this sense, the $n$th raw moment of the conditional probability density \cref{eq:averagep} is defined as follows:
\begin{align}
  %\E(\dlgiven) = \frac{2}{Nz} \sum_{i=1}^{Nz/2} \intinf (\Delta \ell)\cdot p_{\Delta l_i|\bar l_i=\bar \ell}(\Delta \ell)\, d(\Delta \ell)
  \E\left[(\dlgiven)^n\right] \coloneqq \frac{2}{Nz} \sum_{i=1}^{Nz/2} \intinf x^n\, p_{\Delta l_i|\bar l_i=\bar \ell}(x)\, dx\,.
  \label{eq:moments}
  %\Var(\dlgiven) = \frac{2}{Nz} \sum_{i=1}^{Nz/2} \intinf x^2\, p_{\Delta l_i|\bar l_i=\bar \ell}(x)\, dx\\
\end{align}

In the following we characterize the conditional probability density given in \cref{eq:averagep} that completely determines the final distribution of spring lengths given the initial distribution (\cref{eq:plf}).
%ence, we need to derive the \emph{conditional} probability density for the spring length changes $p_{\dl | \lbar=\ellbar}(\dell)$.Since $p_{\lbar}(\ellbar)$ is known, 
%because
%\begin{align}
%p_{\dl | \lbar}(x) = \frac{2}{Nz} \sum_{i=1}^{Nz/2} p_{\Delta l_i | \bar l_i}(x)\,.
%\end{align}
%We start with the unconditional density $p_{\Delta l_i}$. 
Reconsidering \cref{eq:kkt}, we write 
\begin{align}
\dl = \lstar - \lbar = (\mathbf P- \mathbf I)\lbar\eqqcolon \mathbf S \lbar\,.
\label{eq:defS}
\end{align}
\Cref{eq:defS} relates $\dl$ to $\lbar$ and a random matrix $\mathbf S$, both of which vary with the topology of each realization.
%\sm{Introduce difficulties in more detail here?}
It is therefore challenging to obtain $p_{\dlgiven}$ explicitly, especially since the individual $\bar l_i$ are not mutually independent.
Instead, we consider the first two moments of the probability distribution, $\E( \dlgiven )$ and $\vardlgivenl$, and investigate under which conditions $\dl|_{\lbar=\ellbar}$ is approximately normally distributed.
%It is apparent from \cref{eq:dli} that $\mu_{\Delta l_i} = 0$, since $\mu_{\bar l_i}=0$, and hence via \cref{eq:averagedDensity}: $\mu_{\dl}=0$.

%In the following investigations we

In the following we will work with conditional random variables, so we now highlight two important aspects of our generation procedure that will be used extensively.
The first is that each graph cycle, and therefore constraint, contains at least three edges, implying that the edge lengths are pairwise independent as random variables.
%i.e., $p_{\bar l_j |{\bar l_i}}(x) = p_{\bar l_j}(x)$ for all $i,j = 1,\dots,Nz/2$ with $i \neq j$.
The second aspect is that we can fix the abstract graph structure in our generation procedure, leading to $(N,z)$-ensembles with varying winding numbers, but with a constant $\mathbf{S}$-matrix (e.g., \cref{img:schematic}). These fixed-graph-ensembles still contain identically, uniformly distributed random variables $\bar l_j \sim \mathcal U(-0.5,0.5)$.
%, implying $p_{\bar l_j |_{\mathbf{S}}} = p_{\bar l_j}$ for all $j = 1, \dots, Nz/2$. 

\subsection{Conditional mean} % (fold)
\label{sub:Conditional mean}
In this section we compute $\E(\dlgiven)$ for $(N,z)$-ensembles.
We first derive the conditional mean for a fixed-graph-ensemble, i.e., $\E[ (\dlgiven) | \mathbf S ]$, and then generalize the result to $(N,z)$-ensembles.
%Defining the random variable $\boldsymbol X\coloneqq (\dlgiven)$, 
% subsection Conditional mean (end)
\Cref{eq:defS,eq:moments} lead to:
%\begin{align}
%{$\E(\Delta l_i|\bar l_i = \ellbar) = S_{ii} \ellbar$},
%\end{align}
%where $\E(X)$ refers to the expected value of $X$ (just as $\mu_X$),
%and therefore:% \cref{eq:averagedDensity}:
\begin{equation}
  \begin{split}
    \E[(&\dlgiven) | \mathbf S] =
  %\E[ (\dlgiven) | \mathbf S= \fixS ] =
  \frac{2}{Nz}\sum\limits_{i=1}^{Nz/2} \E[ (\Delta l_i|\bar l_i = \ellbar)|\mathbf S]\\%=\boldsymbol{\mathcal S} ]\\
  %=\frac{2}{Nz}\sum\limits_{i=1}^{Nz/2}\E\left[\left(\left(\sum\limits_{j=1}^{Nz/2} S_{ij}\bar l_j\right)|\bar l_i = \ellbar\right)|\mathbf S = \fixS \right]\\
  %=\frac{2}{Nz}\sum\limits_{i=1}^{Nz/2}\E\left[\left(\sum\limits_{j=1}^{Nz/2} S_{ij}\bar l_j + S_{ii} \ellbar\right)|\mathbf S = \fixS \right]\\
  &=\frac{2}{Nz}\sum\limits_{i=1}^{Nz/2}\left(S_{ii} \ellbar+\sum\limits_{j=1,\,j\neq i}^{Nz/2}S_{ij}\underbrace{\E[\bar l_j|\bar l_i = \bar \ell]}_{=\E[\bar l_j]=0} \right) \\
  %=\frac{2}{Nz}\sum\limits_{i,j=1}^{Nz/2}\mathcal \E[ \bar l_j|\bar l_i = \ellbar ] \\
  %= \frac{2}{Nz}\sum\limits_{i=1}^{Nz/2} S_{ii} \ellbar
  &= \frac{2\,\ellbar}{Nz} \tr \mathbf S\,,
  \label{eq:muconditional1}
\end{split}
  %= - \frac{2\, \bar l}{z} \left(1-\frac{1}{N}\right)\,,
\end{equation}
where we used the fact that fixed-graph-ensembles have uniformly distributed edge random variables that are pairwise independent.   %$p_{\bar l_j |_{\bar l_i}} = p_{\bar l_j}$ and $p_{\bar l_j |_{\mathbf{S}}} = p_{\bar l_j}$.}
%\begin{align}
%%\begin{equation}
%  %\begin{split}
%  \E( \dlgiven )
%  %\frac{2}{Nz}\sum%\limits_{i=1}^{Nz/2} \E(\Delta l_i|\bar l_i = \ellbar)\\&=
%  %\frac{2}{Nz}\sum\limits_{i=1}^{Nz/2} S_{ii} \ellbar
%  = \frac{2\,\ellbar}{Nz} \tr \mathbf S = - \frac{2\, \ellbar}{z} \left(1-\frac{1}{N}\right)\,, \label{eq:muconditional}
%%\end{split}
%  %= - \frac{2\, \bar l}{z} \left(1-\frac{1}{N}\right)\,,
%%\end{equation}
%\end{align}
We further make use of our knowledge about the graph's cycle matrix $\mathbf C$ to determine $\tr \mathbf S$. First note that by definition ${\tr \mathbf S = \tr \mathbf P-Nz/2}$ (\cref{eq:defS}).
The projector property of $\mathbf P$ (i.e., ${\mathbf P^2=\mathbf P}$) leads to ${\tr \mathbf P = \dim (\im \mathbf P) = Nz/2 - \dim(\ker \mathbf P)}$, because $\mathbf P$ has eigenvalues $0$ and $1$ only.
Furthermore, $\ker \mathbf P = \ker \mathbf C$, by definition (\cref{eq:kkt}), and hence $\tr \mathbf S = - \dim(\ker \mathbf C)$.
Recall that $\mathbf C$ contains $N(z/2-1)+1$ linearly independent rows corresponding to a set of fundamental cycles of the graph, i.e., $\mathbf C$ has full rank and so $\dim(\ker \mathbf C) = Nz/2 - (N(z/2-1)+1) = N-1$. It follows that
\begin{align}
\tr \mathbf S = 1-N
\label{eq:traceS}
\end{align}
is an invariant of the $(N,z)$-ensemble as it surprisingly only depends on the number of nodes in the graph.
Making use of this invariance together with the general property of the expected value, $\E(X)=\E_Y[\E(X|Y)]$, we combine \cref{eq:muconditional1,eq:traceS} to obtain:
%\begin{align}
%  \E[ (\dlgiven) | \mathbf S ] = - \frac{2\, \ellbar}{z} \left(1-\frac{1}{N}\right)\,.
%  \label{eq:muconditionalS}
%\end{align}
%Using that \cref{eq:muconditionalS} is the same for all $\mathbf S$, i.e., depends on $N$ and $z$ only, and the general property of the expected value, $\E(X)=\E_Y[\E(X|Y)]$, we obtain that
\begin{equation}
  \begin{split}
  \E(\dlgiven) &= \E_{\mathbf S}\big[\E[ (\dlgiven) | \mathbf S ]\big]\\
  &=- \frac{2\, \ellbar}{z} \left(1-\frac{1}{N}\right)\,.
  \end{split}
  \label{eq:muconditional}
\end{equation}
%where the last equality holds due to \cref{eq:traceS}.
\subsection{Conditional variance} % (fold)
\label{sub:Conditional variance}

% subsection Conditional variance (end)
The conditional variance $\vardlgivenl$ remains challenging to express analytically for arbitrary $z$ and $N$.
%\sn{This is because, in general, there are many graphs realizing the same $z$ and $N$, and even a single graph structure allows for different sets of winding numbers (\cref{img:schematic} (c)).}
%To overcome these difficulties for the
To compute the conditional mean, we used the essential fact that \mw{expectation is always additive regardless of the dependencies between the random variables. This is not true for variances.
The variance is only additive if the terms are pairwise independent.}
%By first considering fixed-graph-ensembles we can overcome one challenge (as we will see in a later section to compute VARDELL).
While the edge random variables are pairwise independent, they are in general not conditionally pairwise independent since the remaining two edge random variables of a triangle (cycle with three edges) with one edge length fixed are coupled by the fact that they sum to an integer winding number.
For an $(N,z)$-ensemble, we do not know the abstract graphs, let alone their triangle structures, making the general computation of the conditional variance difficult.
%\so{Knut: This was true for the mean as well; the challenge is more (i) dependencies, and (ii) unknown $\sum_{i \neq j} S_{ij}$.
%Mention triangle stuff? Conditional pairwise independence breaks down as soon as a graph contains triangles; dont know which graphs contain how many triangles.}
%\sout{each having a different---and unknown---cycle structure that changes the dependencies between the edge lengths.}

%var paragraph: there is only one graph, both of which are highly symmetric (edge and cycle transitivity) allowing us to derive variance explcitly.
%in the intermediate there are many graphs with teh same N,z but the variance still shows a smooth.

For two extreme cases, namely the cycle graph ($z=2$, $N>3$) and the complete graph ($z=N-1$, each node connected to every other node), there exists only a single possible graph with a known triangle structure. Both are symmetric (i.e., vertex- and edge-transitive \cite{biggs1993algebraic}).
In particular, edge-transitivity (informally: edges are indistinguishable from each other) allows us to reduce to a single entry in $\lstar$, since $p_{\lstar} = p_{l_i^*}$.
%\so{Informally speaking, edge transitivity says that all edges look the same, implying $p_{\lstar} = p_{l_i^*}$.
The single component $l_i^*$ is given by a weighted sum of identically distributed, but dependent random variables (\cref{eq:kkt}), which we analyze to derive $\vardlgivenl$ explicitly.
%These weights are given by the entries in the $\mathbf S$-matrix, which is shown in \cref{fig:smatrix} for the cycle and complete graph as well as a particular realization of an intermediate regime ensemble.
%Observe the symmetry exposed in the two extreme cases.
%\sm{This dependency poses significant difficulties for expressing the final distribution.
%\sm{To this end, we reduce the number of variables by relaxing the integer cycle constraint to an interval constraint.}
%\sn{For the case of the cycle graph this results in a set of independent random variables, which then leads to $\vardlgivenl=(N-1)/N^2 \Var (\bar l)$}

We first present the conditional variance derivation for these extreme cases and then discuss the more general intermediate-connectivity regime $2<z<N-1$ that contains ensembles of multiple graphs.
The complexity in this regime is highlighted by the intricacies involved in deriving the variance for a fixed graph (e.g., the complete graph), which already requires a special choice of basis to obtain a tractable expression for $(\mathbf C \mathbf C^T)^{-1}$. 

\subsubsection{Cycle graph}
For the cycle graph ($z=2$), there is only one cycle that contains all $N$ edges.
Therefore, the cycle matrix can be written as $\mathbf C = (1,1,\dots,1) \in \mathbb R^{N}$.
It follows that \cref{eq:kkt} simplifies to
\begin{align}
  \lstar = N^{-1} \mathbf C^T\mathbf C \lbar = (g/N) \mathbf I \,,
  %\lstar = \frac{1}{N}\mathbf C^T\mathbf C \lbar = \frac{1}{N} \mathbf J \lbar = \frac{g}{N} \mathbf I \,,
  \label{eq:kkt-cycle}
\end{align}
where $g=\sum_{j=1}^N \bar l_j$ is the winding number of the cycle
%$\mathbf J \in \mathbb R^{N \times N}$ is the matrix of ones,
and $\mathbf I\in \mathbb R^N$ is the vector of ones.
% subsection Conditional probability density (end)
% subsection Variance (end)
We derive the conditional variance $\vardlgivenl = N^{-1} \sum_{i=1}^N \Var( \Delta l_i| \bar l_i= \bar \ell )$ for the cycle graph.
By edge-transitivity $\vardlgivenl = \Var( \Delta l_i| \bar l_i= \bar \ell )$ and by \cref{eq:kkt-cycle}, $\Delta l_i = l^*_i - \bar l_i = g/N - \bar l_i = \sum_{j=1}^N \bar l_j/N - \bar l_i$. % and using conditional pairwise independence yields:
%Now, the variance of a sum is the sum of the variances for pairwise independent random variables.
For the cycle graph, if $N>3$, the conditional edge random variables are pairwise independent, so we compute:
%(The cHowever, for $N=3$ the cycle graph is equal to the complete graph; we hence refer to following section for the derivation.
\begin{align}
  %\sigma_{\Delta l_i| \bar l_i=\epsilon}^2 = \frac{1}{N^2} \mu_{g^2|\bar l_i=\epsilon} - \left( \mu_{g|\bar l_i=\epsilon} \right)
 %\Var[\Delta l_i| \bar l_i=\epsilon] = \frac{1}{N^2}\left[ \E(g^2|\bar l_i=\epsilon) - \E(g|\bar l_i=\epsilon)^2\right]\,.\\
 \Var( \Delta l_i| \bar l_i=\bar\ell ) &= \frac{1}{N^2} \sum_{j=1,\,j\neq i}^N \Var (\bar l_j)\\
 &= \frac{N-1}{N^2} \Var( \bar l )\,,
 \label{eq:var-dlicond}
\end{align}
a constant that is independent of the initial spring length $\bar \ell$, showing that $\vardlgivenl = \meanvar$.
Conditional pairwise independence only holds for $N>3$ because, for $N=3$, if we condition on one length ($\bar l_1=\barell$) the remaining two lengths are dependent via $\bar l_3 = g-\barell - \bar l_2$. 
Therefore, a similar computation for the conditional variance of a general ensemble does not hold, since each graph may contain triangles with conditionally pairwise dependent edges.
%As previously mentioned, more general graphs contain triangles whose edge random variables are then conditionally pairwise dependent, so that a similar computation does not hold.
%where we have used conditional pairwise independence of the initial edge random variables. % $\bar l_j$.
%We notice that
%\begin{align}
%  g|_{l_i = \epsilon} = \epsilon + \sum_{i\neq j} \bar l_j
%  \label{eq:gcycle}
%\end{align}
%and thereby:
%\begin{align}
%  \E(g^2|\bar l_i=\epsilon) = \E [(\epsilon + \sum_{j \neq i} \bar l_j)^2]= \epsilon^2 + (N-1) \Var(\bar l)\,,
%\end{align}
%because all mixed terms are zero; in particular, $\E(\epsilon \bar l_j)=0$, since $\E (\bar l_j) =0$, and $\E(\bar l_m \bar l_n)_{m\neq n} = 0$, because the $\{\bar l_j\}$ are pairwise independent. 
%Note that the latter only holds for $N>3$.
%For $N=3$, the cycle graph is equal to the complete graph. We therefore refer to the section about the complete graph for its variance derivation.
%Analogously it holds that
%\begin{align}
%  \E(g|\bar l_i=\epsilon) = \E [\epsilon + \sum_{j \neq i} \bar l_j]= \epsilon\,.
%\end{align}
%Substituting in \cref{eq:var-dlicond} leads to
%\begin{align}
%  \Var[\Delta l_i| \bar l_i=\epsilon] = \frac{N-1}{N^2} \sigma_{\bar l}^2 = \sigma^2_{\dl | \lbar} \,.
%  \label{eq:vardlcycle}
%\end{align}
\begin{figure*}
  \centering
  \includegraphics{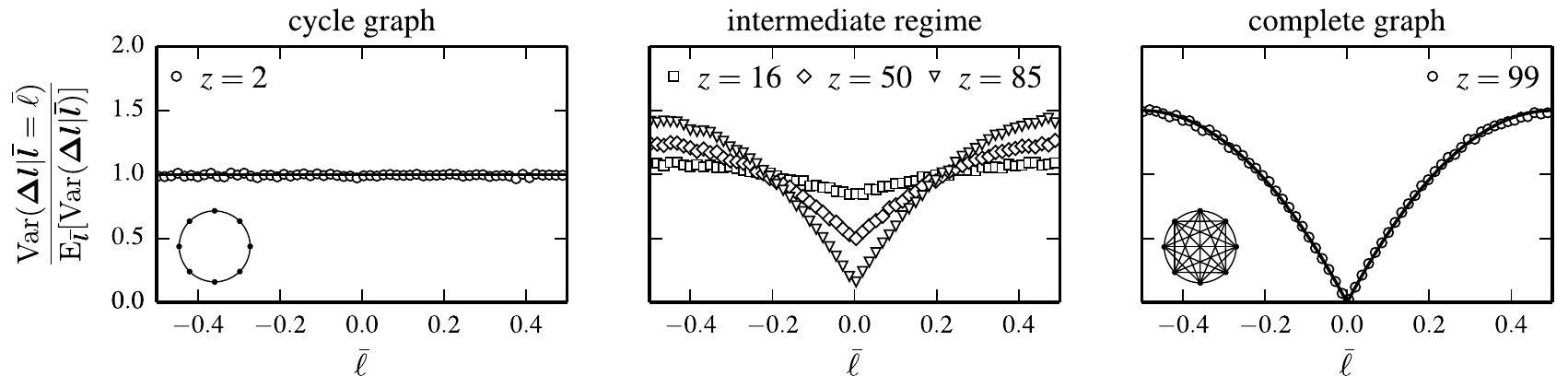}
  \caption{Normalized conditional variance $\vardlgivenl/\meanvar$ as a function of $\bar \ell$ for graphs with $N=100$ and varying $z$ values. For each value of $z$, data points correspond to ensemble averages (repeated simulations) with \num{4.95e6} springs in total. We use local linear regression with \num{3e4} nearest neighbors to estimate the variance for different values of $\bar \ell$. The solid lines correspond to the analytically derived expressions for cycle and complete graph (illustrated in the insets). In the intermediate regime of connectivity, the variance shows a continuous transition between the two extreme cases.}
  \label{fig:conditionalVariance}
\end{figure*}
%The expression in~\cref{eq:var-dlicond} is consistent with Eq.~(12) in the main article for $z=2$.
%\sn{For the case of the cycle graph, an entry in \cref{eq:kkt} simplifies to $\Delta l_i = N^{-1}\sum_{j\neq i} \bar l_j$. Using conditional pairwise independence of the initial edge random variables allows for direct computation of $\vardlgivenl=(N-1)/N^2 \Var (\bar l)$.}
%
%Harnessing this independence requires a non-standard transformation of random variables, which complicates a direct application of the Berry-Esseen Theorem to obtain a quantitative bound on the distance to a normal distrubution.}
%We find that for the cycle graph, $\vardlgivenl=(N-1)/N^2 \Var (\bar l)$, and for the complete graph,

\subsubsection{Complete graph}
\label{sss:complete}
For the case of the complete graph ($z=N-1$), the derivation of the conditional variance is significantly more involved.
In order to obtain manageable algebraic expressions, {one} needs to carefully choose the cycle basis (i.e., spanning tree). %In particular we restrict to an edge that is contai
This choice of basis leads to a tractable expression for $(\mathbf C \mathbf C^T)^{-1}$, which can then be applied to reformulate the problem in terms of conditionally independent winding number random variables.
%analyze the dependencies of the edge lengths. 
%This allows for a reformulation in terms of conditionally independent winding number random variables that we use to compute the 
%$\vardlgivenl = (N-2)/N^2 (|\ellbar| - \ellbar^2)$.}
\begin{figure}
  \centering
  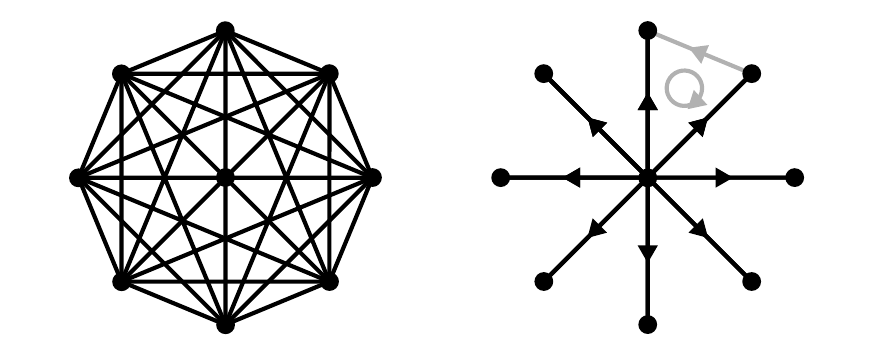
  \caption{(a) The complete graph ($z=N-1$) for $N=9$ vertices, here shown as undirected graph for clarity. (b) A spanning tree of a directed version of the complete graph in (a). The chosen spanning tree is based at the vertex at the center and, from there, reaches out to all $N-1$ other vertices. The edge orientations of the spanning tree edges (black) are chosen such that they have opposite orientation with respect to a connecting cycle (see, e.g., the cycle formed by the gray edge).\label{fig:spanning-tree-complete}}
\end{figure}
%For the case of the complete graph ($z=N-1$) we work with a particular cycle basis by

%\paragraph{Spectral analysis}
We choose a spanning tree as shown in \cref{fig:spanning-tree-complete}.
For the following derivation, we label the edges such that the first $N-1$ edges correspond to the edges of the spanning tree.
The other $m=N(N-1)/2-(N-1)$ edges are the ones that are added to the spanning tree to construct the fundamental cycles (here: all triangles).
We order the cycles in $\mathbf C$ according to these edges and decompose the cycle matrix into two parts:
%and choose aligned cycle orientations.
\begin{align}
  \mathbf C = (\underbrace{\mathbf A}_{N-1} | \underbrace{\mathbf I}_{N(N-1)/2-(N-1)})\,,
  \label{eq:Cdecomp}
\end{align}
where $\mathbf I$ is the identity matrix.

Our first result is that for spanning tree edges $\boldsymbol l^*_\text{st}\coloneqq \{l^*_i\}_{i=1}^{N-1}$ we have that 
\begin{align}
\boldsymbol l^*_{\text{st}} =
N^{-1} (\C^T \C)_\text{st} \cdot \lbar=
N^{-1}
\begin{pmatrix}
\A^T \A & \A^T
\end{pmatrix}
\cdot \lbar\,,
  \label{eq:ctc}
\end{align}
where $(\C^T \C)_\text{st}$ corresponds to the first $N-1$ rows of $\C^T \C$. The importance of this result is that the symmetries of the complete graph allow it to extend to all edges.
Indeed, each vertex of the complete graph defines a spanning tree as shown in \cref{fig:spanning-tree-complete}; therefore every edge can be seen as such a spanning tree edge. Independence of the solution of the choice of the cycle basis, and therefore spanning tree, implies that the following derivations hold for all edges in the graph.
 
Since, by \cref{eq:kkt,eq:Cdecomp},
\begin{align*}
  (\lstar)_\text{st} = 
\left(
  \mathbf A^T (\mathbf A \mathbf A^T + \mathbf I)^{-1} \mathbf A \quad \mathbf A^T(\mathbf A \mathbf A^T + \mathbf I)^{-1} \right)\cdot \lbar\,,
%\label{eq:kkt-decomp}
\end{align*}
proving \cref{eq:ctc} is equivalent to showing that
%
%It follows that $(\mathbf C \mathbf C^T)^{-1} = (\mathbf A \mathbf A^T + \mathbf I)^{-1}$. 
%By substituting \cref{eq:Cdecomp} into \cref{eq:kkt} we \mw{find \sout{get} }
%\begin{align}
%  \lstar =
%\begin{pmatrix}
%  \mathbf A^T (\mathbf A \mathbf A^T + \mathbf I)^{-1} \mathbf A & \mathbf A^T(\mathbf A \mathbf A^T + \mathbf I)^{-1} \\
%  (\mathbf A \mathbf A^T + \mathbf I)^{-1} \mathbf A & (\mathbf A \mathbf A^T + \mathbf I)^{-1} 
%\end{pmatrix} \cdot \lbar\,.
%\label{eq:kkt-decomp}
%\end{align}
%%We show that for the spanning tree edges, $\boldsymbol l^*_\text{st}\coloneqq \{l^*_i\}_{i=1}^{N-1}$, \cref{eq:kkt-decomp} simplifies, just as for the cycle graph, to \cref{eq:kkt-cycle}.
%Noticing that
%\begin{align}
% \C^T \C = 
%  \begin{pmatrix}
%  \A^T \A & \A^T \\ \A & \mathbf I
%  \end{pmatrix}\,,
%\end{align}
%\mw{MAKE LEMMA --- STRANGE FORMULATION ! we} aim to derive that for the spanning tree edges, $\boldsymbol l^*_\text{st}\coloneqq \{l^*_i\}_{i=1}^{N-1}$, \cref{eq:kkt-decomp} simplifies to:
%\begin{align}
%\boldsymbol l^*_{\text{st}} =
%N^{-1} (\C^T \C)_\text{st} \cdot \lbar=
%N^{-1}
%\begin{pmatrix}
%\A^T \A & \A^T
%\end{pmatrix}
%\cdot \lbar\,,
%  \label{eq:ctc}
%\end{align}
%\mw{where the \sout{whose} } middle term resembles the cycle graph setting (\cref{eq:kkt-cycle}).
%\begin{lem}
%The matrix $A$ satisfies the following identities:
\begin{align}
  %\item $\mathbf A^T (\mathbf A \mathbf A^T + \mathbf I)^{-1} \mathbf A = \frac{1}{N} \A^T \A\,,$
  \mathbf A^T(\mathbf A \mathbf A^T + \mathbf I)^{-1} = N^{-1}\A^T \nonumber\\
  \Leftrightarrow \;  \left[(N-1)\mathbf I - \mathbf A^T \mathbf A\right] \mathbf A^T = 0\,.
  \label{eq:rightAT}
\end{align}
%\end{lem}
%\begin{proof}
% section Complete graph (end)
%The first identity follows from the second by multiplication by $\mathbf A$ on the right.
%We now prove the second identity by first showing thatThe second identity is equivalent to showing that
%  \begin{align}
%    (\A \A^T)^2 = (\A \A^T)(N-1)\,.
%    \label{eq:AATsquared}
%  \end{align}
%    \label{lem:AATsquared}
%\end{lem}
%\begin{proof}
We can construct $\A^T \A \in \mathbb R^{(N-1)\times(N-1)}$ explicitly: The $ij$th entry counts the number of cycles that are shared by the edges $i$ and $j$---with a contribution of $1$ if the edges have the same orientation with respect to the cycle, and $-1$ otherwise.
As can be seen in \cref{fig:spanning-tree-complete}, two spanning tree edges only share one cycle with opposite orientations, hence $(A^T A)_{ij}=-1$ for $i\neq j$.
Each edge is itself part of $N-2$ cycles, so $(A^T A)_{ij}=N-2$ for $i=j$:
\begin{align}
  \A^T \A  = 
  \begin{cases}
    N-2, \quad &\text{when} \; i=j \\
    -1, \quad &\text{when} \; i\neq j 
  \end{cases}
  = (N-1) \mathbf I - \mathbf J\,,
  \label{eq:ATA}
\end{align}
where $\mathbf J \in \mathbb R^{(N-1)\times(N-1)}$ is the matrix with ones everywhere.
Substitution into \cref{eq:rightAT} yields:
\begin{align}
  \mathbf J \mathbf A^T = 0\,.
  \label{eq:ja}
\end{align} 
%\mw{It thus remains to show that $\im(\mathbf A^T) \in \ker(\mathbf J)$.}
\Cref{eq:ja} holds true since each column of $\mathbf A^T$ contains two nonzero entries, $1$ and $-1$, due to all fundamental cycles (triangles) involving two spanning tree edges with opposite orientations (\cref{fig:spanning-tree-complete}).

We have therefore shown that for the spanning tree edges of the complete graph, the following relation holds:
\begin{align}
  \boldsymbol{l^*_{\text{st}}} 
%\frac{1}{N}
%\begin{pmatrix}
%\A^T \A & \A^T
%\end{pmatrix}
%\cdot \lbar
= N^{-1} (\C^T\C)_{\text{st}} \cdot \lbar = N^{-1} (\C^T \boldsymbol g)_\text{st}\,,
\label{eq:spanningedges}
\end{align}
where $(\C^T \boldsymbol g)_\text{st}$ is the vector of the first $N-1$ entries of $\C^T \boldsymbol g$, and $\boldsymbol g = \C \lbar$ is the vector of winding numbers (\cref{eq:totalEnergy}).
\Cref{eq:spanningedges} allows us to change perspective to winding number random variables.
For a particular edge in the spanning tree, we compute $l^*_\text{st} = \frac{1}{N}\sum_{j=1}^{N-2}g_j$, and therefore, $\Delta l_\text{st} = \frac{1}{N}\sum_{j=1}^{N-2}g_j - \bar l_\text{st}\,,$
%\begin{align}
  %\Delta l_\text{st} = l^*_\text{st} - \bar l_\text{st} = \frac{1}{N}\sum\limits_{j=1}^{N-2}g_j - \bar l_\text{st}\,,
  %\label{eq:lstarsumg}
%\end{align}
since the edge is contained in exactly $N-2$ fundamental cycles, which we have assumed correspond to the first $N-2$ entries in the $\boldsymbol g$ vector, and edge and cycle orientations are aligned.
For the conditional random variable $\Delta l_\text{st}|_{\bar l_{\text{st}}=\barell}$, it follows that: 
%this task into analyzing the winding numbers $g_j|_{\barlst=\barell}$ of the fundamental cycles:
\begin{align}
\Delta l_\text{st}|_{\bar l_{\text{st}}=\barell} = \frac{1}{N}\sum\limits_{j=1}^{N-2}g_j|_{\barlst=\barell} - \barell\,.
  \label{eq:lstarsumgcond}
\end{align}

Observe that the $g_j|_{\barlst=\barell}$ are independent random variables since their only potential dependence, their common edge, is conditioned out.
%Therefore, knowing the variance of the $g_j|_{\barlst=\barell}$ would yield the variance of $\Delta l_\text{st}|_{\bar l_{\text{st}}=\barell}$.
Each winding number $g_j|_{\barlst=\barell}$ corresponds to a fundamental cycle that is a triangle (\cref{fig:spanning-tree-complete}), i.e., involves only three edges of which one is fixed.
Therefore we cannot use conditional pairwise independence of edge lengths as in the case of the cycle graph to compute the variance.
Instead, we derive the winding number distribution of a triangle explicitly.

The initial edge lengths are distributed as $\bar l_j \sim \mathcal U(-0.5,0.5)$, so the winding numbers can only attain three values $\{-1,0,1\}$. 
In particular, $g_j|_{\bar l_\text{st}=\barell} = \barell + \bar l_{j_1} +\bar l_{j_2}$, where we choose positive signs since the lengths are distributed symmetrically around zero.
We compute the probability of $g_j|_{\bar l_\text{st}=\barell}$ attaining the value zero:
\begin{equation}
  \begin{split}
  P(g_j|_{\bar l_\text{st}=\barell}=0) &= P(\bar l_{j_2} \in [-\barell - 0.5, -\barell + 0.5])\\
  &= \int \limits_{-\barell - 0.5}^{-\barell + 0.5} \chi_{[-0.5,0.5]}(x) \, dx = 1 - |\barell|\,,
  \label{eq:gdist1}
\end{split}
\end{equation}
where $\chi_{[-0.5,0.5]}(\cdot)$ is the characteristic function on the interval $[-0.5,0.5]$.
The remaining probability is assigned to either $g_j|_{\bar l_\text{st}=\barell}=1$ or $g_j|_{\bar l_\text{st}=\barell}=-1$ depending on whether the given $\barell$ is positive or negative:
\begin{align}
  P(g_j|_{\bar l_\text{st}=\barell\geq 0}= 1) = P(g_j|_{\bar l_\text{st}=\barell\leq 0}= -1) = |\barell|\,.
  \label{eq:gdist2}
\end{align}
%We conclude that $g_j|_{\barlst=\epsilon}$ is essentially a \textit{Bernoulli} random variable with success probability $|\epsilon|$.

%To measure how far $\Delta l_\text{st}|_{\bar l_{\text{st}}=\barell}$ is from being normally distributed for finite $N$ we look at the standardized random variable
%\begin{align}
%  Y_{N-2} \coloneqq \frac{\Delta l_\text{st}|_{\bar l_{\text{st}}=\barell}-\E(\Delta l_\text{st}|{\bar l_{\text{st}}=\barell})}{\sqrt{\Var(\Delta l_\text{st}|{\bar l_{\text{st}}=\barell})}}
%\end{align}
%and compare its cdf to the one of the standard normal.
%\paragraph{Variance computation}
Using \cref{eq:lstarsumgcond}, conditional independence of the $g_j|_{\bar l_\text{st}=\barell}$, and their probability distribution, \cref{eq:gdist1,eq:gdist2}, we have:
\begin{align}
  %\E(\Delta l_\text{st}|\bar l_{\text{st}}=\barell) &= \frac{N-2}{N}\E(g_j|\bar l_\text{st}=\barell) - \barell = \frac{N-2}{N}\barell - \barell\,, \\
  \Var(\Delta l_\text{st}|{\bar l_{\text{st}}=\barell}) &= \frac{N-2}{N^2}\Var(g_j|\bar l_\text{st}=\barell)\\ 
  &=  \frac{N-2}{N^2} (|\barell|-\barell^2) \,.
  \label{eq:varcomplete}
\end{align}
In contrast to the cycle graph (\cref{eq:var-dlicond}), the conditional variance $\vardlgivenl = \Var(\Delta l_\text{st}|{\bar l_{\text{st}}=\barell})$ for the complete graph (\cref{eq:varcomplete}) depends on the initial spring length $\bar \ell$, as is shown in \cref{fig:conditionalVariance}.
\subsubsection{Intermediate-connectivity regime}
\sn{For the intermediate-connectivity regime, $2<z<N-1$, a tractable expression for $(\mathbf C \mathbf C^T)^{-1}$, as for the complete graph, remains elusive; however, numerical data suggest that the variance exhibits a continuous transition between the two extremes (\cref{fig:conditionalVariance}).}
%\sn{despite consiting of varying topologies, both in winding numbers and graphs.}
%\sm{We will discuss under which circumstances the conditional variance is approximately constant, i.e., $\vardlgivenl\approx\meanvar$.}
We also observe that the conditional variance is approximately constant given that $z \ll N$. This is the most relevant case for biological networks where typically $z\lesssim4$.
For $z\ll N$, we may thus approximate $\vardlgivenl\approx\meanvar$\sn{, which we now derive.}

%We perform a thorough analysis by considerin
The \emph{law of total variance} \cite{weiss2006course} \sn{states}:
\begin{align}
\E_{\lbar}[\Var(\dl\,|\,\lbar)] = \Var( \dl ) - \Var_{\lbar}[\E( \dl\,|\,\lbar)]\,.
\label{eq:totalvar}
\end{align}
%Each cycle constraint involves at least three elements of $\lbar$ leading to \emph{pairwise independence} between the $\bar l_j$.
%This allows for calculating the variance of $\Delta l_i$, needed for $\Var( \dl )$, via \cref{eq:defS}:
%\begin{align}
%  \Var( \Delta l_i ) = \sum_{j=1}^{Nz/2} S_{ij}^2 \Var( \bar l_j ) = \Var( \bar l )\sum_{j=1}^{Nz/2} S_{ij}^2 \,,
%\end{align}
%where the second equality follows from all $\bar l_j$ being equally distributed.
%Again, we use that $\mathbf P^2=\mathbf P$, hence $\mathbf S^2 = -\mathbf S$, and therefore:
%\begin{align}
%  \sum_{j=1}^{Nz/2} S_{ij}^2 = -S_{ii} \; \Rightarrow \;  \Var( \Delta l_i ) = -S_{ii}\,\Var( \bar l ) \,.
%\end{align}
We first compute $\Var( \dl )$, again, by initially fixing $\mathbf S$ and considering $\Var( \dl | \mathbf S )$. 
%%\so{Unlike for conditional edge lengths, the unconditioned edge lengths are all pairwise independent. 
%Each cycle constraint involves at least three elements of $\lbar$, leading to pairwise independence between the $\bar l_j$. ANDY: CHANGE TO PAIRWISE INDEPENDENCE}
With \cref{eq:defS,eq:moments} we compute
\begin{align*}
  %\Var( \Delta l_i ) = \sum_{j=1}^{Nz/2} S_{ij}^2 \Var( \bar l_j ) = \Var( \bar l )\sum_{j=1}^{Nz/2} S_{ij}^2 \,,
  \Var[ \dl | \mathbf S ] = \frac{2}{Nz}\sum_{i=1}^{Nz/2} \Var(\Delta l_i | \mathbf S) =\frac{2 \Var(\bar l)}{Nz}\sum_{i,j=1}^{Nz/2} S_{ij}^2\,,
  %\\ \sum_{j=1}^{Nz/2} S_{ij}^2 \Var( \bar l_j ) = \Var( \bar l )\sum_{j=1}^{Nz/2} S_{ij}^2 \,, 
\end{align*}
%\so{where the second equality follows from $\bar l_j |_{\mathbf{S}} = \bar l_j$ and all $\bar l_j$ being pairwise independent and identically distributed.}
where the second equality follows from fixed-graph ensembles having uniformly distributed edge random variables that are pairwise independent.   %$p_{\bar l_j |_{\bar l_i}} = p_{\bar l_j}$ and $p_{\bar l_j |_{\mathbf{S}}} = p_{\bar l_j}$.}
Again, we use that $\mathbf P^2=\mathbf P$, hence $\mathbf S^2 = -\mathbf S$, and therefore $\sum_{j=1}^{Nz/2} S_{ij}^2 = -S_{ii}$.
%\begin{align}
%  \sum_{j=1}^{Nz/2} S_{ij}^2 = -S_{ii} \; \Rightarrow \;  \Var( \Delta l_i ) = -S_{ii}\,\Var( \bar l ) \,.
%\end{align}
Insertion into the equation above yields:
\begin{align}
  \frac{\Var( \dl | \mathbf S )}{\Var(\bar l)} =-\frac{2\tr \mathbf S}{Nz} = \frac{2}{z} \left(1-\frac{1}{N}\right)\,,
\end{align}
where the second equality is due to \cref{eq:traceS}.
%This result is an invariant of the $(N,z)$-ensemble.
Application of the law of total variance gives:
\begin{equation}
  \begin{split}
  \Var(\dl) &= \E_{\mathbf S}[\Var(\dl | \mathbf S)] + \Var_{\mathbf S}[\underbrace{\E(\dl | \mathbf S)}_{=0}] \\
  &= \frac{2}{z}\left(1-\frac{1}{N}\right) \Var(\bar l)\,,
  \label{eq:sigmadl2}
\end{split}
\end{equation}
where the second term in the sum vanishes using \cref{eq:defS} since $\E(\bar l_i)=0$ (analogous to the computation in \cref{eq:muconditional1}).
From \cref{eq:muconditional} we can use
%\begin{align}
$\Var_{\lbar}[\E( \dl\,|\,\lbar)] = {(2/z(1-1/N))^2 \Var(\bar l)}$
  %$\sigma^2_{\mu_{\Delta l\,|\,\bar l}} = 1/n^2 \sigma^2_{\bar l}$
  %\label{eqvaredl}
%\end{align}
and therefore find by substituting into \cref{eq:totalvar}:
%\begin{align}
%  \E_{\bar l}(\Var[\Delta l\,|\,\bar l]) =\Var (\bar l)\left(\frac{1}{n} - \frac{1}{n^2} \right) \,.
%  \label{eq:meanvar}
%\end{align}
\begin{align}
  %\left\langle\sigma^2_{\dl|\lbar}\right\rangle_{p_{\lbar}} &=\frac{2}{z}\left(1-\frac{1}{N}\right)\left[1 - \frac{2}{z}\left(1-\frac{1}{N}\right)\right] \sigma^2_{\bar l}\,.
  \frac{\E_{\lbar} [\vardlgiven]}{\varlbar} &=\frac{2}{z}\left(1-\frac{1}{N}\right)\left[1 - \frac{2}{z}\left(1-\frac{1}{N}\right)\right] \,.
  %&= \frac{1}{n^2} (n-1) \, \sigma_{\bar l}^2 \,.
  \label{eq:meanvar}
\end{align}

\subsection{Normality} % (fold)
\label{sub:Normality}

% subsection Normality (end)
If $\dl|_{\lbar=\ellbar}$ were normally distributed, having estimates for mean and variance (\cref{eq:muconditional,eq:meanvar}) would be sufficient to fully characterize $p_{\dlgiven}$.
Indeed, for the two extremes, cycle and complete graph, we can prove that $\dl|_{\lbar=\ellbar} $ is \emph{normally distributed} in the limit $N \to \infty$, with a rate of convergence proportional to $(N-2)^{-1/2}$.
%\footnote{For the complete graph $\bar l$ needs to be bounded away from zero because the error bound $\sim (|\bar l| - {\bar l}^2)^{-1/2}$}.

This result might look like a direct application of the classical \emph{central limit theorem}.
However, since the edge lengths are not independent as random variables, more sophisticated techniques are required to represent the solution in terms of a suitable set of \sn{mutually} independent random variables.
In contrast to situations in time series analysis \cite{brockwell2009time}, where independence holds beyond a certain time window, in our case the cycle constraints prohibit localization of dependencies.
\sn{To deal with this problem, we reduce the number of variables by relaxing each integer cycle constraint to an interval constraint.}
%%, for which we again rely on the relaxation of integer to interval constraints.
Harnessing the resulting independence then requires a non-standard transformation of random variables, which complicates a direct application of the Berry-Esseen theorem \cite{berry1941accuracy,esseen1942liapounoff} (a deviation-bound version of the central-limit theorem) to obtain a quantitative bound on the distance to a normal distribution.

%We proceed by mimicking the structure of the conditional variance section.
\mw{The rest of this section is split into three parts.
We begin by proving the results for the cycle and complete graph, and then investigate the intermediate-connectivity regime.
%Throughout, notice how the proofs may not directly generalize to the Intermediate-connectivity regime where ensembles are no longer symmetric and contain varying graph structures.
%Throughout, note how the proofs of the extreme cases do not generalize in the intermediate-connectivity regime, where ensembles have varying graph structure and lack symmetry.}
Throughout, note how the intricacies of the proofs of the extreme cases are further complicated in the intermediate-connectivity regime, where ensembles have varying graph structure and lack symmetry.}
\subsubsection{Cycle graph}
For the cycle graph ($z=2$), we prove that $\dlc$ is normally distributed in the limit $N \to \infty$.
The key idea is a relaxation of the integer constraint to an interval constraint.
Using $\Delta l_i =  g/N - \bar l_i = \sum_{j=1}^N \bar l_j/N - \bar l_i$ (\cref{eq:kkt-cycle}) we introduce the standardized ($\E( Y_N ) =0$, $\Var( Y_N )=1$) random variable
\begin{align}
  Y_{N} &\coloneqq \frac{\dlc-\E(\dlC)}{\sqrt{\Var(\dlC)}}\\
  &= \frac{\gc/N - \barell - (\barell/N -\barell)}{\sqrt{\frac{N-1}{N^2} \varlbar}}\\
  &= \frac{\gc - \barell}{\sqrt{(N-1)\varlbar}}\,,
\end{align}
and compare its cdf $F_{Y_N}(x)$ to that of the standard normal $\Phi_{0,1}(x)$.
We find that
%Defining $a \coloneqq \sigma_{\bar l}\sqrt{N-1}$ it holds that
\begin{align}
  F_{Y_N}(x) &= \Prob(Y_N \leq x)\\
  &= \Prob \left(g|_{\bar l_i=\barell} \leq x\,\sqrt{(N-1)\varlbar}  + \barell\right)\\
  &= \sum_{k = -\infty}^{\left\lfloor x\,\sqrt{(N-1)\varlbar} + \barell \right\rfloor} \Prob(g|_{\bar l_i=\barell} = k)\,,
  \label{eq:Fcycle1}
\end{align}
where we have used the fact that $g \in \mathbb Z$ and $\lfloor \cdot \rfloor$ denotes the floor operator.
Using $g|_{l_i = \barell} = \barell + \sum_{j=1, \,j\neq i}^N \bar l_j$
%\begin{align}
%g|_{l_i = \barell} = \barell + \sum_{j=1, \,j\neq i}^N \bar l_j
%\label{eq:gcond}
%\end{align}
we can formalize the integer relaxation by expressing the probability of the conditional winding number as follows: 
\begin{align}
  \Prob(g|_{\bar l_i=\barell}=k) = \int_{k-\barell-0.5}^{k-\barell+0.5} \nu_{N-2}(t)\, dt\,,
  \label{eq:probg}
\end{align}
where $\nu_{N-2}(t)$ corresponds to the probability density of the sum of $N-2$ uniformly distributed independent random variables on the interval $[-0.5,0.5]$. We call this random variable $U_{N-2}$.
There are only $N-2$ independent random variables because one of the $N$ lengths is fixed to $\barell$ and another one is determined to make sure that an integer winding number is attained for $g$.
Substituting \cref{eq:probg} into \cref{eq:Fcycle1} and expressing $\left\lfloor x\,\sqrt{(N-1)\varlbar}+ \barell\right\rfloor = x\,\sqrt{(N-1)\varlbar} + \barell - \delta(x)$, with the random variable $\delta(x) \in [0,1)$, leads to:
\begin{align*}
  F_{Y_N}(x) &= \sum_{k = -\infty}^{x\,\sqrt{(N-1)\varlbar} + \barell -\delta(x)} \int_{k-\barell - 0.5}^{k -\barell + 0.5} \nu_{N-2}(t)\, dt \\
  &= \int_{-\infty}^{x\,\sqrt{(N-1)\varlbar} + 0.5 -\delta(x)} \nu_{N-2}(t)\, dt \\&= F_{U_{N-2}}\left(x\,\sqrt{(N-1)\varlbar} + 0.5 - \delta(x)\right)\,.
  %= F_{\frac{U - 0.5 + \delta}{a}}(x)\,.
\end{align*}

We are interested in the distance to the cdf $\Phi_{0,1}(x)$ of the standard normal distribution.
To calculate this distance, this we perform a change of variables which results in a standardized sum of uniforms $U_{N-2}/(\sigmaU)$, where $\sigma_{\bar l}=\sqrt{\Var{(\bar l)}}$.
However, this causes the cdf of our random variable and the standard normal cdf to have different arguments.
We therefore split the computation into two steps: one that measures the distance to a shifted standard normal cdf, and the other that measures the deviations introduced by this shift.
With the shorthand notation $\newx \coloneqq x\,\sqrt{(N-1)\varlbar} + 0.5 - \delta(x)$, the described procedure corresponds to the following computation: 
%\begin{widetext}
%\begin{align*}
%  %\sup_{x\in\mathbb R}
%  &|F_{Y_N}(x) - \Phi_{0,1}(x)|=\left|F_{U_{N-2}}\left(  \sqrt{(N-1)\varlbar}x + 0.5 - \delta(x)\right) - \Phi_{0,1}(x)\right|\nonumber \\[1ex]
%  &=\left|F_{\frac{U_{N-2}}{\sqrt{(N-2)\varlbar}}}\left(\frac{\sqrt{(N-1)\varlbar}x + 0.5 - \delta(x)}{\sqrt{(N-2)\varlbar}}\right) - \Phi_{0,1}(x)\right|\nonumber\\[1ex]
%  %\begin{split}
%  &\leq\underbrace{\left|F_{\frac{U_{N-2}}{\sqrt{(N-2)\varlbar}}}\left(\frac{\sqrt{(N-1)\varlbar}x + 0.5 - \delta(x)}{\sqrt{(N-2)\varlbar}}\right) - \Phi_{0,1}\left(\frac{\sqrt{(N-1)\varlbar}x + 0.5 - \delta(x)}{\sqrt{(N-2)\varlbar}}\right)\right|}_{\eqqcolon I(x)}
%  %\label{eq:firstterm}
%  \\[1ex]
%  &+\underbrace{\left|\Phi_{0,1}\left(\frac{\sqrt{(N-1)\varlbar}x + 0.5 - \delta(x)}{\sqrt{(N-2)\varlbar}}\right)-\Phi_{0,1}(x)\right|}_{\eqqcolon II(x)}\,.
%  %\label{eq:secondterm}
%%\end{split}
%  %&= |F_{\frac{U}{a}}(x-(\delta - 0.5)/a) - \Phi_{0,1}((x-(\delta - 0.5)/a))| +|\Phi_{\frac{\delta - 0.5}{a},1}(x) - \Phi_{0,1}(x)|
%\end{align*}
%\end{widetext}
\begin{align*}
  %\sup_{x\in\mathbb R}
  &|F_{Y_N}(x) - \Phi_{0,1}(x)|=\left|F_{U_{N-2}}\left( \newx \right) - \Phi_{0,1}(x)\right|\nonumber \\[1ex]
  &=\left|F_{\frac{U_{N-2}}{\sigmaU}}\left(\frac{\newx}{\sigmaU}\right) - \Phi_{0,1}(x)\right|\nonumber\\[1ex]
  %\begin{split}
  &\leq\underbrace{\left|F_{\frac{U_{N-2}}{\sigmaU}}\left(\frac{\newx}{\sigmaU}\right) - \Phi_{0,1}\left(\frac{\newx}{\sigmaU}\right)\right|}_{\eqqcolon I(x)}
  %\label{eq:firstterm}
  \\[1ex]
  &+\underbrace{\left|\Phi_{0,1}\left(\frac{\newx}{\sigmaU}\right)-\Phi_{0,1}(x)\right|}_{\eqqcolon II(x)}\,.
  %\label{eq:secondterm}
%\end{split}
  %&= |F_{\frac{U}{a}}(x-(\delta - 0.5)/a) - \Phi_{0,1}((x-(\delta - 0.5)/a))| +|\Phi_{\frac{\delta - 0.5}{a},1}(x) - \Phi_{0,1}(x)|
\end{align*}
$I(x)$ can be bounded using the Berry-Esseen theorem. Bounding $II(x)$ requires a detailed case analysis (see~\cref{sec:app1} for details). We arrive at:
%\begin{widetext}
\begin{align}
   &\supr |F_{Y_N}(x) - \Phi_{0,1}(x)| \leq I(x) + II(x)\\
   &\leq\frac{12^{3/2} \,C}{32\sqrt{N-2}}  +\frac{1}{\sqrt{2 \pi (N-2)\varlbar}}\\
   &= \frac{1}{\sqrt{N-2}}\left(\frac{12^{3/2} \,C}{32}  +\frac{1}{\sqrt{2 \pi\varlbar}}\right)\,.
   \label{eq:boundcycle}
\end{align}
%\end{widetext}
Therefore, the cdf of $\dlc$ converges to a normal distribution with the rate $(N-2)^{-1/2}$, independent of $\barell$.
%Analogous to the definition in \cref{eq:cumcdf}, $F_{\dlgiven}(x)\coloneqq N^{-1} \sum_{i=1}^N F_{\Delta l_i|\bar l_i=\barell}\,(x)$.
Since we showed that $F_{\Delta l_i|\bar l_i=\barell}$ is independent of the edge $i$, \cref{eq:cumcdf} implies $F_{\dlgiven}\,(x)=F_{\Delta l_i|\bar l_i=\barell}\,(x)$, and therefore $F_{\dlgiven}$ converges to a normal distribution as well.
\subsubsection{Complete graph}

For the complete graph, our proof of normality relies on the reduction to spanning tree edges as outlined in the conditional variance section.
In particular, this allows us to write the cdf in terms of winding number random variables that are all triangles that share a common edge.
Conditioning on this edge then yields independence, not of the length variables, but of these winding number random variables, which allows us to apply the Berry-Esseen theorem.

To measure how far $\Delta l_\text{st}|_{\bar l_{\text{st}}=\barell}$ is from being normally distributed for finite $N$, we look at the standardized random variable
\begin{align}
  Y_{N-2} \coloneqq \frac{\Delta l_\text{st}|_{\bar l_{\text{st}}=\barell}-\E(\Delta l_\text{st}|{\bar l_{\text{st}}=\barell})}{\sqrt{\Var(\Delta l_\text{st}|{\bar l_{\text{st}}=\barell})}}
\end{align}
and compare its cdf to the one of the standard normal.
Using \cref{eq:lstarsumgcond} and the probability distribution of $g_j|_{\bar l_\text{st}=\barell}$, \cref{eq:gdist1,eq:gdist2}, we obtain:
\begin{align}
  \E(\Delta l_\text{st}|\bar l_{\text{st}}=\barell) &= \frac{N-2}{N}\E(g_j|\bar l_\text{st}=\barell) - \barell \\
  &= \frac{(N-2)\barell}{N}- \barell\,, \\
  \Var(\Delta l_\text{st}|{\bar l_{\text{st}}=\barell}) &= \frac{N-2}{N^2}\Var(g_j|\bar l_\text{st}=\barell) \\&=  \frac{N-2}{N^2} (|\barell|-\barell^2)\,,
  \label{eq:varcomplete2}
\end{align}
and therefore
\begin{align}
  Y_{N-2} &= \frac{\frac{1}{N}\sum_{j=1}^{N-2} g_j|_{\bar l_\text{st}=\barell} - \frac{(N-2)\barell}{N}}{\sqrt{\frac{N-2}{N^2}\Var(g_j|\bar l_\text{st}=\barell)}} \\
  &= \frac{\sum_{j=1}^{N-2}(g_j|_{\bar l_\text{st}=\barell} -\barell)}{\sqrt{\sum_{j=1}^{N-2} \Var(g_j|_{\bar l_\text{st}=\barell}-\barell)}}\,.
\end{align}
All $\{g_j|_{\bar l_\text{st}=\barell}\}_{j=1}^{N-2}$ are independent since the corresponding cycles only share one edge, which is the one that we condition on.
We can thus apply the Berry-Esseen theorem (\cref{thm:berryesseen}) to show that for $N\geq3$,
\begin{align}
  \supr |F_{Y_{N-2}}(x)-\Phi_{0,1}(x)| \leq \frac{C \rho}{\sigma^3 \sqrt{N-2}}\,,
\end{align}
with $C < 0.4748$, $\rho = \E(\left|g_j|_{\bar l_\text{st}=\barell}-\barell\right|^3)$, and $\sigma^2 = \Var(g_j|_{\bar l_\text{st}=\barell}-\barell)=|\barell|-\barell^2$. Computing $\rho = (|\barell|-\barell^2)(\barell^2+(1-|\barell|)^2)$ via \cref{eq:gdist1,eq:gdist2} we arrive at
\begin{align}
  \supr |F_{Y_{N-2}}(x)-\Phi_{0,1}(x)| \leq \frac{C}{\sqrt{N-2}} \frac{\barell^2+(1-|\barell|)^2}{\sqrt{|\barell|-\barell^2}}\,,
  \label{eq:boundcomplete}
\end{align}
for $|\barell| > 0$. This proves convergence of the cdf of $\dlc$ to a normal distribution with the rate $(N-2)^{-1/2}$.
For $\barell = 0$, $\dlc \sim \delta_0$ (Dirac delta distribution around zero), it can only attain the value zero because $P(g_j|_{\bar l_\text{st}=\barell}=0) = 1$. 

Note the $\barell$-dependence in \cref{eq:boundcomplete}, which is in stark contrast to the $\barell$-independent bound for the cycle graph (\cref{eq:boundcycle}).
For the complete graph, the approximation with a normal distribution becomes worse as $\barell$ approaches zero.% as can be seen in~\cref{fig:conditional}.
%\begin{figure}
%  %\showthe\columnwidth
%  \centering
%  \includegraphics{img/conditional-densities-supplement-N-100-newlabels.pdf}
%  \caption{Probability density for the conditional length changes $p_{\dlgiven}$ for graphs with $N=100$ and $z=50$ (left), $z=99$ (right). The symbols correspond to ensemble averages (repeated simulations) with \num{4.95e6} springs in total. Solid lines correspond to best fit Gaussians. For small values of $\bar l$---i.e., smaller variance---we observe larger deviations from normality, as suggested by \cref{eq:boundcomplete} for the complete graph (here: $z=99$). Even for less connected networks (left graph, $z=50$), a similar effect can be observed.\label{fig:conditional}} 
%\end{figure}
%
%Observe also that \cref{eq:varcomplete} proves that for the complete graph:
%\begin{align*}
%\vardlgivenl = \frac{2}{Nz} \sum_{i=1}^{Nz/2} \Var( \Delta l_i|\bar l_i = \barell ) =  \frac{N-2}{N^2}(|\barell| - \barell^2)\,,
%\end{align*}
%as presented in the main article.
\subsubsection{Intermediate-connectivity regime}
\begin{figure}
  \centering
  \includegraphics{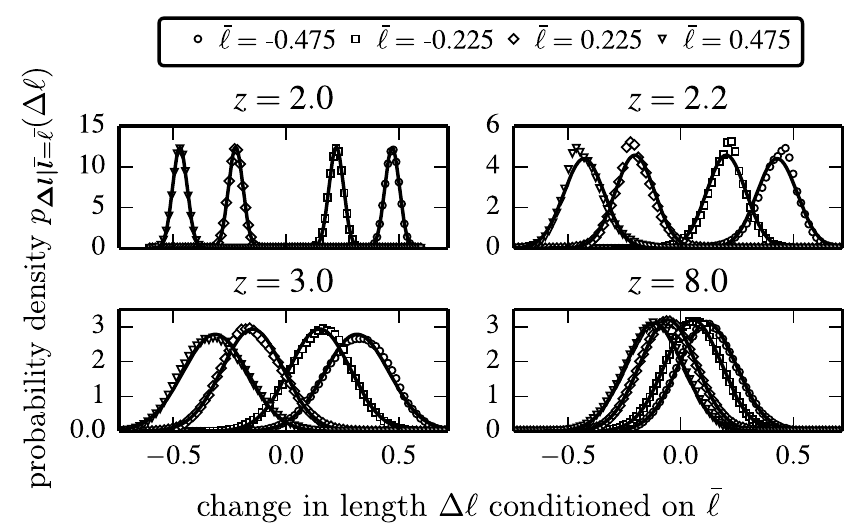}
  \caption{Conditional probability density $p_{\dlgiven}\,(\Delta \ell)$ for spring networks with $N=100$ and varying $z$, conditioned on different $\bar \ell$ values. For each value of $z$, data points correspond to ensemble averages (repeated simulations) with \num{4.95e6} springs in total. Solid lines correspond to best-fit normal distributions. The cycle graph ($z=2$) is close to being normally distributed---as proven for $N\to\infty$. Whereas for $z=2.2$, there are still deviations from a normal distribution, for $z=3$ and larger, the densities rapidly approach a normal distribution.}
  \label{fig:pdlconditional}
\end{figure}
Recall that \sn{in the intermediate-connectivity regime, $2<z<N-1$, the $(N,z)$-ensembles contain graphs with varying cycle structures making a similar analysis significantly more challenging.}
\sn{In simulations}, \sn{however}, we observe that $\dl|_{\lbar=\ellbar}$ is approximately normally distributed if $z$ is sufficiently large (\cref{fig:pdlconditional}).

\subsection{Density approximation} % (fold)
\label{sub:Density approximation}

\begin{figure}
  \centering
  \includegraphics{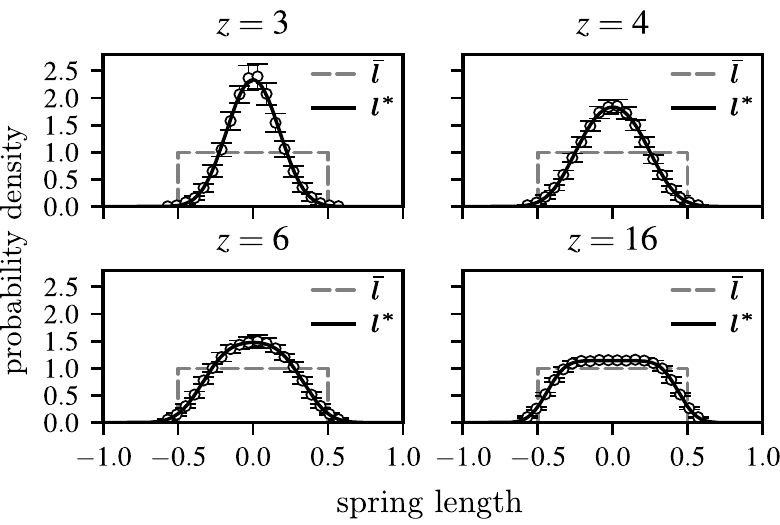}
  %\includegraphics{length-distribution-various-n-small}
  %\caption{Comparison between the \emph{expected} final spring length distribution \cref{eq:meanfield} (solid red curve) and simulations (grey histogram) for \emph{individual} regular spring networks with $N=1000$ and $z=\{3,4,6,16\}$.}
  \caption{Probability density $p_{\lstar}(\ell^*)$ for the final spring lengths for networks with $N=1000$ and varying $z$. Solid black lines show the analytic expression for $p_{\lstar}(\ell^*)$ (\cref{eq:meanfield}); data points correspond to averages over 50 simulations. The error bars correspond to the standard deviation. \sn{For comparison, we show the initial uniform spring length distribution $p_{\lbar}(\ellbar)$ as a gray dashed line.}}
  \label{fig:lengthdistribution}
\end{figure}

% subsection Density approximation (end)
Our empirical observations and theoretical discussion above justify the following approximation for $3\leq z\ll N$:
\begin{align}
  %\dl_{\lbar=\ellbar} \sim \mathcal N\left(\mu_{\dlgiven},\left\langle\sigma^2_{\dl|\lbar}\right\rangle_{p_{\lbar}}\right)\,,
  \dl|_{\lbar=\ellbar} \sim \mathcal N\Big[\E( \dlgiven ),E_{\lbar}[\vardlgiven]\Big]\,,
  \label{eq:pdlconditional}
\end{align}
with the expressions for $\E( \dlgiven )$ and $E_{\lbar}[\vardlgiven]$ given in \cref{eq:muconditional,eq:meanvar}.
%\cref{fig:conditional} shows that the conditional variances measured in the simulations fluctuate around the expected value uniformly.
%We use \cref{eq:meanvar} as an estimate for $\sigma_{\dl | \lbar}^2$.
%\begin{figure}
%  \centering
%  \includegraphics{conditional-heat}
%  \caption{2D histogram showing the joint distribution $p_{\Delta l, \bar l}$ for an average over 20 (?) simulations with $n=5$, $N=1000$. The black crosses refer to conditional means $\mu_{\Delta l | \bar l}$. One can clearly see the dependence $\mu_{\Delta l | \bar l} = -\bar l/n$ (solid black line).}
%  \label{fig:conditionalheat}
%\end{figure}
%\begin{figure}
%  \centering
%  \includegraphics{conditional-variance-n5}
%  \caption{Conditional variance $\sigma_{\Delta l | \bar l}$ as  a function of $\bar l$ for a simulated system with $N=1000$, $n=5$ (grey solid line). The red solid line shows the theoretical value for the average variance $\E_{\bar l}[ \sigma^2_{\Delta l|\bar l}]$ (see \cref{eq:meanvar}).}
%  \label{fig:conditionalvariance}
%\end{figure}
%\cref{fig:conditionalheat} shows this density as function of $\bar l$ for a uniform distribution of initial spring lengths, averaged over several simulated systems.
Using \cref{eq:plf,eq:pdlconditional}, we obtain an explicit representation for the final length distribution $p_{\lstar}(\ellstar)$ in mechanical equilibrium (\cref{sec:bigequation}). 
%We apply this approach to our simulations by choosing $p_{\lbar}$ to be uniform; it follows that $\sigma_{\bar l}^2 = 1/12$.
%and therefore from \cref{eq:meanvar}, $\sigma_{\Delta l | \bar l}^2 \simeq 1/12(1/n-1/n^2)$.
%We obtain an explicit expression for $p_{\lstar}(\ellstar)$ under the assumption
%We obtain:
%  \begin{align}
%  p_{\lstar}(z) =&  \intinf \mathcal U(-0.5,0.5)(a) \cdot \mathcal N (- a/n, \sigma_{\Delta l | \bar l}^2)(z-a) \,da \nonumber \\
%  =&  \frac{1}{\sqrt{2\pi \sigma_{\Delta l | \bar l}^2}} \int\limits_{-0.5}^{0.5}\exp\left[-\left(\frac{z-(1-1/n)a}{\sqrt{2\sigma_{\Delta l | \bar l}^2}}\right)^2\right]\,da \nonumber \\
%  %&-\erf\left.\left(\frac{z-(1-1/n)/2}{\sqrt{2}/12(1/n-1/n^2)}\right)\right]\,.\\
%  \simeq& \frac{1}{2(1-1/n)}\left[\erf\left(\frac{z+(1-1/n)/2}{\sqrt{2(1/n-1/n^2)/12}}\right)\right.\\
%  &-\left.\erf\left(\frac{z-(1-1/n)/2}{\sqrt{2(1/n-1/n^2)/12}}\right)\right]\,. \nonumber
%\end{align}
In \cref{fig:lengthdistribution} we compare this analytical expression to ensembles of simulated networks and observe excellent agreement. 

% section Mean field theory (end)
\section{Comparison to a mean-field approach}
In order to evaluate the significance of our graph-theoretical analysis we compare it to a mean-field (mf) approach which neglects all topological features other than the local degree of connectivity.
%Before investigating when the conditional probability density $p_{\dlgiven}\,(\Delta \ell)$ resembles a normal distribution,
%In particular, we compare our exact results for the conditional mean \cref{eq:muconditional} and unconditional variance \cref{eq:sigmadl2} to those arising from a mean-field approach that .
%We compare \cref{eq:muconditional} to a \emph{mean-field} (mf) approach, 
%for a regular network, where each node in the network has the same degree and
%where each node is displaced as if all other nodes in the network were fixed.
\begin{figure}
  \centering
  %\showthe\columnwidth
  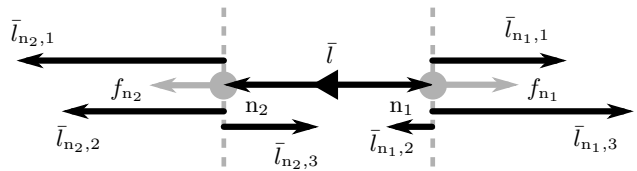
  \caption{Mean-field approach (here: $z=4$): The edge $\bar l$ connects two nodes $(\mathrm{n_1},\mathrm{n_2})$.
  %Upon relaxation, these nodes are displaced independently according to the
  The initial node forces (gray arrows) are given by $f_{\mathrm n_1}= \sum_{i=1}^{z-1} \bar l_{\mathrm n_1,i}+\bar l$ and $f_{\mathrm n_2}= \sum_{i=1}^{z-1} \bar l_{\mathrm n_2,i} - \bar l$. The spring $\bar l$ contributes to the forces with different signs because its length is measured from $\mathrm{n_1}$ to $\mathrm{n_2}$ (depicted by the black triangle). While displacing an individual node by $u_\mathrm{n_k} = z^{-1}f_\mathrm{n_k}$ introduces force balance at that node, this approach neglects that nodes/edges are coupled, i.e., force balance has to be established at all nodes simultaneously, as in \cref{eq:totalEnergy}.}
  \label{fig:mean-field}
\end{figure}
In contrast to the graph-theoretical model, where $z$ refers to the \emph{average} degree of a node, the mean-field approach assumes that each node is connected to \emph{exactly} $z$ other nodes.
Moreover, the node displacement $u_\text{node}$ during relaxation is calculated as if all other nodes in the network were fixed. 
Therefore $u_\text{node} = f_\text{node}/z$,
%\begin{align}
%\end{align}
where $f_\text{node}=\sum_{i=1}^z \bar l_i$ is the initial force acting on the node via the springs attached to it.
The displacement $\Delta l$ of a spring is given by the difference of the displacements of the two nodes that are connected by this edge:
\begin{equation}
  \begin{split}
  \Delta l &= u_{\mathrm n_2} - u_{\mathrm n_1} = z^{-1}(f_{\mathrm n_2}-f_{\mathrm n_1})\\
  &=\frac{1}{z} \left(-2 \bar l  + \sum_{i=1}^{z-1} \bar l_{\mathrm n_2,i}-\sum_{i=1}^{z-1} \bar l_{\mathrm n_1,i} \right)\,,
  \label{eq:deltal}
\end{split}
\end{equation}
where we have taken into account that the two nodes share one spring, namely $\bar l$ (\cref{fig:mean-field}).
All springs are assumed to be independent identically distributed random variables with mean zero and variance $\varlbar$.

%\subsection{Conditional mean}
For the conditional mean, we have with \cref{eq:deltal}:
\begin{align}
  \E(\dlgiven)|_\text{mf} = -\frac{2 \barell}{z}\,.
\end{align}
%as presented in the main article.
%according to the initially acting force transmitted by the springs attached to it (see~supplementary material for detailed derivation).
%, i.e., $\Delta x_i = f_i/z$; all other nodes in the network are treated as fixed.}
%In this case ${\E( \dlgiven )|_{\text{mf}} = - 2 \ellbar/z }$ (see~SM for details);
The mean-field result agrees with the exact solution \cref{eq:muconditional} in the limit $N \to \infty$, i.e., there is no significant difference for large node numbers.
In contrast, we will show at the end of this section that for the variance, the mean-field solution differs substantially from the exact result, even in the limit $N \to \infty$.

When considering normality of $\dl|_{\lbar=\ellbar}$, the mean-field approach allows us to directly apply the Berry-Esseen theorem (\cref{thm:berryesseen}) because all edges are treated as independent.
Defining the normalized random variable 
\begin{align}
  Y_{2(z-1)} :=& \frac{\Delta l|_{\bar l=\barell}-\E(\Delta l|{\bar l}=\barell)}{\sqrt{\Var(\Delta l|{\bar l=\barell})}}\\
  =&  \frac{\sum_{i=1}^{2(z-1)} \bar l_i}{\sqrt{2(z-1) \Var(\bar l)}}\,,
\end{align}
the theorem implies:
\begin{align}
  \supr |F_{Y_{2(z-1)}}(x)-\Phi_{0,1}(x)| \leq \frac{12^{3/2} \,C}{32\sqrt{2(z-1)}}\,.
\end{align}
The mean-field approach yields convergence to a normal distribution with a rate proportional to $(z-1)^{-1/2}$.
While this result agrees with the rate of convergence we proved for the complete graph, it is in stark contrast to what we proved for the cycle graph case ($z=2$), for which we showed convergence to a normal distribution even though $z$ is constant. % for the cases $z=2$ and $z=N-1$ with a rate proportional to $(N-2)^{-1/2}$, i.e., dependent on the number of nodes, not the connectivity.
%In particular, for $z=2$, a mean-field approach does not recover normality due to negligence of global edge coupling.
In the intermediate-connectivity regime, both the mean-field as well as our graph-theoretical approach suggest that  $\dl|_{\lbar=\ellbar}$ can be approximated by a normal distribution.
To complete the evaluation of our approach, it is therefore critical to also compare the second moments, i.e., the variances of the mean-field and graph-theoretical approach.
%Our results therefore suggest approximating the conditional distribution in the intermediate regime with a normal distribution.
%This approach suggests that a mean-field approach may not recover normality due to negligence of global edge coupling.

\begin{figure}
  \centering
  \includegraphics{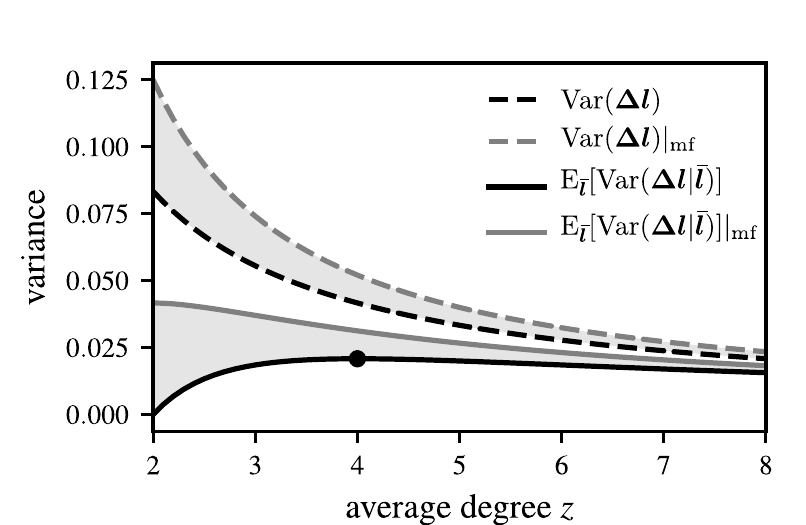}
  \caption{Comparison of graph-theoretical (black) and mean-field (gray) variances as a function of average degree $z$. Shown are the unconditional variance $\vardl$ as well as the expected conditional variance $\meanvar$ as derived in the text.
  The graph-theoretically derived expected variance exhibits a maximum at $z=4$ (filled circle), while the corresponding mean-field expected variance monotonically decreases from its value at $z=2$.}
  \label{img:mfcomparisonvar}
\end{figure}

%\subsection{Conditional variance}
For the unconditional variance, we obtain using \cref{eq:deltal}:
\begin{align}
  \Var( \dl )|_\text{mf}&= z^{-2} \left( 4 \varlbar + 2(z-1)  \varlbar \right) \\&= \frac{2}{z} \left( 1+\frac{1}{z} \right) \varlbar\,.
  \label{eq:vardl}
\end{align}
%$\Var(\dl)|_{\text{mf}}= 2 /z (1+1/z))\varlbar$.
Clearly, this expression does not agree with the exact graph-theoretical value \cref{eq:sigmadl2}, even in the limit $N\to\infty$. 
A mean-field approach assumes the conditional variance is constant and therefore equal to its expected value $\E_{\lbar} [\vardlgiven]|_\text{mf}=\vardlgivenl|_\text{mf}={2/z(1-1/z) \Var(\bar l)}$.
For the cycle graph, we proved the conditional variance is indeed constant (\cref{eq:var-dlicond}). 
However, we showed that the other extreme, the complete graph, exhibits non-constant conditional variance (\cref{eq:varcomplete}).
For $(N,z)$-ensembles in the intermediate-connectivity regime, we observe a continuous transition between the two extremes (\cref{img:mfcomparisonvar}).
\mw{Therefore, for the biological regime ($z \lesssim 4$), we approximated the conditional variance with its constant expected value $\E_{\lbar}[\vardlgiven]$ (\cref{eq:meanvar}).}
%For sparsely connected networks (small values of $z$), we also approximated with constant variance based on our proof that the cycle graph exhibits constant conditional variance (\cref{eq:var-dlicond}).
However, it is exactly the regime $z\lesssim 4$ where the graph-theoretically derived expected conditional variance $\E_{\lbar}[\vardlgiven]$ and the mean-field quantity $\E_{\lbar} [\vardlgiven]|_\text{mf}$ exhibit the largest discrepancy (\cref{img:mfcomparisonvar}).
%Hence, our results are indeed a significant step towards a \ldots of fiber networks beyond classical mean-field descriptions.
%Similarly, the expected conditional variance $\E_{\lbar} [\vardlgiven]|_\text{mf}=\vardlgivenl|_\text{mf}={2/z(1-1/z) \Var(\bar l)}$ does not agree with \cref{eq:meanvar}.
%Especially for sparsely connected networks (small values of $z$), there are significant deviations, independent of the number of nodes in the network (\cref{img:mfcomparisonvar}).
%Furthermore, we can prove for the complete graph that the condtional variance is not constant (\cref{eq:varcomplete}) and we observe a smooth transition the conditional variance functions   can prove it for the complete graph .
%The intermediate seems to be a smooth transition between these two extremes.
%Moreover, the mean-field approach assumes constant conditional variance, which does not agree with our observations for larger $z$ (\cref{fig:conditionalVariance}).
%We observe dependence of $\vardlgivenl$ on $\bar \ell$ (\cref{fig:conditionalVariance}), which we can prove for the complete graph \cref{eq:varcomplete}.

\section{Discussion and conclusions} % (fold)
%\label{sec:Discussion and conclusions}
%We present a probabilistic theory of force distributions in a random disordered spring network.
%In a network with initially unbalanced forces at nodes, the resulting change in length of individual filaments can be excellently approximated as a conditional distribution on the initial force in the filament.
%We obtain the conditional distribution as normally distributed with the mean and variance fully determined by the network parameters.
%For a given initial force distribution, using the conditional distribution of change in lengths we obtain the force distribution in the network once it attains mechanical equilibrium.
%\sn{we need some message as well: we find.... OR...surprisingly our findings suggest....OR..our approach can be useful in .....}
In conclusion, we have presented a probabilistic theory of force distributions in one-dimensional random spring networks on a circle.
Here we have regarded networks with initially unbalanced forces that relax into mechanical equilibrium.
\sn{When drawing the analogy to a biological network, our approach, which focuses on the relaxation of the system after non-equilibrium starting conditions, is equivalent to assuming a separation of time scales where internal or external non-equilibrium processes slowly create forces in the network that rapidly equilibrate.}

%We have constructed the final distribution of spring lengths via the distribution of spring length \emph{changes} given their initial lengths.
We developed a \emph{graph-theoretical} approach that allows us to exactly compute mean and expected variance of the distribution of length changes conditioned on an initial configuration. 
For the two extreme cases, the \emph{cycle graph} and the \emph{complete graph}, we could prove convergence of \sn{this distribution} to a \emph{normal distribution}.
A systematic analytical treatment of the---less symmetric---intermediate-connectivity regime is more demanding and not provided here.
\sn{However, our results suggest an approximation that shows excellent agreement with simulation for the biologically relevant regime of connectivity, $3\leq z\ll N$}.

It is straightforward to generalize the approach we present here to higher spatial dimensions $d$ if the probability densities $p_{\lbar_k}$ for the components of the initial spring vectors are independent.
In that case, due to the linearity of spring forces with extension, the optimization problem decouples into the spatial components.
The probability density for the final spring vectors then is simply given as the product of the one-dimensional results:
\begin{align}
  p_{\lstar}(\boldsymbol{\ell^*}) = \prod\limits_{k=1}^d p_{\lstar_k}(\ell^*_k)\,.
  \label{eq:plfhigherdimension}
\end{align}
Hence, our results carry over to two- and three-dimensional networks, which are more commonly studied in practice and are of biological and physiological relevance.

Interestingly, a classical \emph{mean-field} approach \emph{fails} to correctly reproduce the mean and the variance of the relevant distributions. The error is particularly pronounced for the---biologically most relevant---regime of low degrees of connectivity, and does not vanish in the limit of infinite node number.
%\sm{The small $z$ regime is the physically most relevant since typcial disordered filamentous networks have a degree of connectivity in the range $2<z \leq 4$ (refs).}
Our work demonstrates that \emph{network topology}---here manifested as \emph{cycle constraints}---is crucial for the correct determination of force distributions in an elastic spring network.

%We consider this study a starting point for future research, in several ways.
%On one hand, the setting of \emph{elastic networks} offers several natural extensions, e.g., \emph{deformed} networks, network \emph{dynamics}, or \emph{nonlinear} springs.
%: (i) Can we apply the proposed concepts to networks that are being \emph{deformed} (quasistatically)?%, i.e., is it possible to derive bulk elastic properties based on the full graph representation of the network?
%Also, is it possible to relate \emph{stress heterogeneities}, like force chains, to the topology of the network?
%(ii) Is it possible to transfer  
%On the other hand, the concepts presented in this work might also prove useful for network theories, other than elastic networks.
%Furthermore, our simulation approach guarantees exact conservation of cycle contractabilities and therefore provides the appropriate framework for simulations of periodic spring type networks --- or more general, networks on \ldots
This opens the door for future research on the role of network topology in more complex elastic networks, e.g., in the presence of dynamics, spring nonlinearities or rupture.
Moreover, the mixture of probabilistic and graph-theoretical techniques may prove useful for other types of network theories.
% section Discussion and conclusions (end)
\begin{acknowledgments}
  The authors would like to thank Friedrich Bös, Alexander Hartmann, and Fabian Telchow for fruitful discussions. Funding from the Deutsche Forschungsgemeinschaft (DFG) within the collaborative research center SFB 755, project A3, is gratefully acknowledged.
C.F.S was additionally supported by a European Research Council Advanced Grant PF7 ERC-2013-AdG, Project 340528.
\end{acknowledgments}
\begin{widetext}
\begin{appendix}
\section{Independence of the choice of cycle basis} % (fold)
\label{sec:Independence of the solution on the choice of the cycle basis}
% section Independence of the solution on the choice of the cycle basis (end)
%The general solution to the quadratic programming problem was given as
%\begin{align}
%  \lstar =\underbrace{\mathbf C^T(\mathbf C\mathbf C^T)^{-1}\mathbf C}_{\eqqcolon \mathbf P} \lbar \quad \Leftrightarrow \quad \dl =\underbrace{(\mathbf P - \mathbf I)}_{\coloneqq \mathbf S} \lbar\,,
%  \label{eq:kkt}
%\end{align}
%with $\mathbf C \in  \mathbb R^{m \times Nz/2}$ the so-called cycle matrix.
%The idea is to represent $\mathbf C$ in a particular \textit{cycle basis} that leads to a concise representation of \cref{eq:kkt}.
%We first show that the solution is independent of the choice of cycle basis.
A change of cycle basis corresponds to the transformation
\begin{align}
  \mathbf{\tilde C} = \mathbf Q \C \mathbf \Per^{-1}\,,
  \label{eq:changeofbasis}
\end{align}
where $\mathbf Q \in \mathrm{GL}(m)$ is an arbitrary change of basis matrix for the cycle space, \mw{and $\Per \in \mathrm{O}(Nz/2)$ is a permutation matrix that corresponds to relabeling the edges of the graph.}
The independence of the solution of the cycle matrix means that, given the change of basis in \cref{eq:changeofbasis} and the solution (\cref{eq:kkt})
  \begin{align*}
  \boldsymbol{\tilde l^*} = \CT^T (\CT \CT^T)^{-1} \CT \boldsymbol{\tilde{\bar l}}
\end{align*} 
  to the transformed problem, 
  \begin{align}
\boldsymbol{\tilde l^*} = \Per\, \lstar\,.
    %\boldsymbol{\tilde l^*} = \CT^T (\CT \CT^T)^{-1} \CT \boldsymbol{\tilde{\bar l}} \quad \Leftrightarrow \quad \lstar =\mathbf C^T(\mathbf C\mathbf C^T)^{-1}\mathbf C \lbar\,.
  \end{align}
The above relation can be shown by direct computation:
  %\begin{widetext}
  \begin{align*}
    \CT^T (\CT \CT^T)^{-1} \CT 
    &= (\Q \C \Per^{-1})^T (\Q\C\Per^{-1}(\Q \C \Per^{-1})^T)^{-1} (\Q \C\Per^{-1})
    = \Per^{-T}\C^T \Q^T (\Q\C\Per^{-1}\Per^{-T} \C^T \Q^T)^{-1} (\Q \C\Per^{-1})\\
    &= \Per^{-T}\C^T \Q^T \Q^{-T}(\C \C^T)^{-1} \Q^{-1} \Q \C\Per^{-1}
    = \Per\C^T(\C \C^T)^{-1} \C\Per^{-1}\,,
\end{align*}
  %\end{widetext}
and therefore
\begin{align*}
  \boldsymbol{\tilde l^*} &=  \Per\C^T(\C \C^T)^{-1} \C\Per^{-1}\boldsymbol{\tilde {\bar l}}
  = \Per \C^T(\C \C^T)^{-1} \C \lbar 
  = \Per  \lstar\,.
\end{align*}
  \section{Upper bounds for $\mathbf{I(x)}$ and $\mathbf{II(x)}$}
  \label{sec:app1}
  For a uniform upper bound on the first term $I(x)$, we can apply the Berry-Esseen theorem, which is stated as follows \cite{athreya2006measure}.
\begin{thm}[Berry-Esseen]
  \label{thm:berryesseen}
  Let $X_1, X_2, \cdots$ be independent identically distributed (iid) random variables with $E(X_1)=0$, $E(X_1^2)=\sigma^2>0$, $E(|X_1|^3)=\rho < \infty$. Also, let 
  \begin{align*}
    S_n = \frac{X_1 + X_2 + \cdots + X_n}{\sqrt{n}\sigma}
  \end{align*}
  be the normalized $n$-th partial sum. Denote $F_n$ the cdf of $S_n$, and $\Phi_{0,1}$ the cdf of the standard normal distribution.
 % For the sake of convenience, denote 
 % \begin{align}
 %   \vec \sigma = (\sigma_1, \cdots,\sigma_n), \; \vec \rho = (\rho_1, \cdots ,\rho_n)\,.
 % \end{align}
 Then there exists a positive constant $C<0.4785$ \cite{Tyurin2010} such that
  \begin{align}
    \sup_{x\in \mathbb R} |F_n(x) - \Phi_{0,1}(x)| \leq \frac{C \rho}{\sigma^3 \sqrt n}\,.
    %\text{where} \quad \psi_0 = \psi_0(\vec \sigma, \vec \rho) = \left(\sum_{i=1}^n \sigma_i^2\right)^{-3/2} \cdot \sum_{i=1}^n \rho_i\,.
  \end{align}
\end{thm}
By recalling that $U_{N-2}$ is the sum of $N-2$ independent uniformly distributed random variables on the interval $[-1/2,1/2]$, i.e., with variance $\varlbar = \sigma^2_{\bar l}=1/12$, third absolute moment $\rho_{\bar l} = 1/32$, and mean $\E(\bar l)=0$, we have that $U_{N-2}/(\sigma_{\bar l} \sqrt{N-2})$ is a normalized $n$-th partial sum. % as in \cref{thm:berryesseen}.
The Berry-Esseen theorem therefore implies:
\begin{align}
  \sup_{x \in \mathbb R} I(x) = \sup_{x \in \mathbb R} \left|F_{\frac{U_{N-2}}{\sigma_{\bar l}\sqrt{(N-2)}}}(x) - \Phi_{0,1}(x)\right| 
    \leq \frac{12^{3/2} \,C}{32\sqrt{N-2}}\,.
  \label{eq:firsttermlimit}
\end{align}

An upper bound for the second term $II(x)$ can be found as well.
We write:
%\begin{widetext}
\begin{align}
  II(x) &= \left|\Phi_{0,1}\left(\frac{x\,\sqrt{(N-1)\varlbar} + 0.5 - \delta(x)}{\sqrt{(N-2)\varlbar}}\right)-\Phi_{0,1}(x)\right|\\
  &= \left|\Phi_{0,1}(\alpha x + \beta )-\Phi_{0,1}(x)\right| = \left|\Phi_{0,1}(y)-\Phi_{0,1}(x)\right|\,,
\end{align}
%\end{widetext}
with $\alpha = \sqrt{\frac{N-1}{N-2}}$, $\beta = \frac{0.5-\delta(x)}{\sqrt{(N-2)\varlbar}}$, and $y=\alpha x + \beta$.
There are six cases that need to be distinguished: $(x<0<y)$, $(y<0<x)$, $(x<y<0)$, $(y<x<0)$, $(0<x<y)$, $(0<y<x)$.

For $(y<0<x)$, the following holds:
%\begin{widetext}
\begin{align}
  \left|\Phi_{0,1}(\alpha x + \beta)-\Phi_{0,1}(x)\right| &\leq [({1-\alpha})x-\beta] \,\sup_{x \in \mathbb R}  \Phi'_{0,1}(x)\leq -\frac{\beta}{\sqrt{2 \pi}} \\ &< \frac{1}{2\sqrt{2 \pi (N-2)\varlbar}}\,,
  \label{eq:case1}
\end{align}
%\end{widetext}
where we have used that $1-\alpha <0$ and $\delta(x) \in [0,1)$.
Analogously, the same bound holds for the case $(x<0<y)$.
For the other cases, we can make use of the convexity (concavity) of $\Phi_{0,1}(x)$ for $x<0$ ($x>0$).

For $(x<y<0)$, we have:
%\begin{widetext}
\begin{align}
  \left|\Phi_{0,1}(\alpha x + \beta)-\Phi_{0,1}(x)\right| &\leq [({\alpha-1})x+\beta] \,\Phi'_{0,1}(\alpha x + \beta) \\
  &=[(\alpha-1)x+\beta] (2\pi)^{-1/2} e^{-(\alpha x + \beta)^2/2 }\\
 &\leq \frac{\beta}{\sqrt{2 \pi}}
  < \frac{1}{2\sqrt{2 \pi (N-2)\varlbar}}\,,
  \label{eq:case2}
\end{align}
%\end{widetext}
and analogously the same for $(0<y<x)$.

Finally, for $(0<x<y)$: 
%\begin{widetext}
\begin{align}
  \left|\Phi_{0,1}(\alpha x + \beta)-\Phi_{0,1}(x)\right| &\leq [(\alpha-1)x+\beta] \,\Phi'_{0,1}(x)\\ 
   &=[(\alpha-1)x+\beta] (\sqrt{2\pi})^{-1/2} e^{-x^2/2 }\\
  &\leq \frac{\beta}{\sqrt{2 \pi}} +  \frac{\alpha-1}{\sqrt{2\pi}}  x e^{-x^2/2}
  \leq \frac{\beta}{\sqrt{2 \pi}} + \frac{\alpha -1}{\sqrt{2\pi e}}\\
  &= \frac{\beta}{\sqrt{2 \pi}}+ \frac{1}{\sqrt{2\pi e}}\frac{\sqrt{N-1}-\sqrt{N-2}}{ \sqrt{N-2}}\\
  &\leq \frac{1}{2\sqrt{2 \pi (N-2)\varlbar}}+ \frac{1}{\sqrt{2\pi e}}\frac{1}{2(N-2)} \label{eq:root}\\
  &\leq \frac{2}{2 \sqrt{2 \pi (N-2)\varlbar}}= \frac{1}{\sqrt{2 \pi (N-2)\varlbar}}\label{eq:combineterms}\,,
\end{align}
%\end{widetext}
where we used the concavity of $\sqrt{x}$ in \cref{eq:root} and $N \geq \varlbar/e + 2$ in \cref{eq:combineterms}.
Analogously, the same bound holds for the last remaining case $(y<x<0)$.
Taking the maximum bound of all cases \cref{eq:case1,eq:case2,eq:combineterms} we obtain
\begin{align}
  \supr II(x) \leq \frac{1}{\sqrt{ 2\pi (N-2)\varlbar}}\,.
  \label{eq:II}
\end{align}

\section{Analytical expression for the final length distribution} % (fold)
By combining \cref{eq:plf} with the normal approximation for $\dl|_{\lbar=\ellbar}$ (\cref{eq:pdlconditional}) we obtain:
\begin{align}
  \begin{split}
  %p_{\lstar}(x) \simeq&  \intinf \mathcal U(-0.5,0.5)(a) \cdot \mathcal N (\mu_{\dl | \lbar}, \langle\sigma^2_{\dl|\lbar}\rangle)(x-a) \,da \nonumber 
  p_{\lstar}(\ellstar) =&  \intinf p_{\lbar}\,(\ellbar) \cdot p_{\dlgiven}\,(\ellstar-\ellbar) \,d\ellbar 
  %=  \frac{1}{\sqrt{2\pi \langle\sigma^2_{\dl|\lbar}\rangle}} \int\limits_{-0.5}^{0.5}\exp\left[-\left(\frac{z-a-\mu_{\dl | \lbar}(a)}{\sqrt{2\langle\sigma^2_{\dl|\lbar}\rangle}}\right)^2\right]\,da \nonumber \\
  \simeq  \frac{1}{\sqrt{2\pi \meanvar}} \int\limits_{-0.5}^{0.5}\exp\left[-\left(\frac{\ellstar-(1-\frac{2}{z}(1-\frac{1}{N}))\ellbar}{\sqrt{2 \meanvar}}\right)^2\right]\,d\ellbar \\
  %&-\erf\left.\left(\frac{z-(1-1/n)/2}{\sqrt{2}/12(1/n-1/n^2)}\right)\right]\,.\\
  =& \frac{1}{2(1-\frac{2}{z}(1-\frac{1}{N}))}\left[\erf\left(\frac{\ellstar+(1-\frac{2}{z}(1-\frac{1}{N}))/2}{\sqrt{2\meanvar}}\right)
  -\erf\left(\frac{\ellstar-(1-\frac{2}{z}(1-\frac{1}{N}))/2}{\sqrt{2\meanvar}}\right)\right]\,. 
\end{split}
  \label{eq:meanfield}
\end{align}
This expression is compared to simulated data in \cref{fig:lengthdistribution}.
\label{sec:bigequation}
\end{appendix}
\end{widetext}

%\footnotesize{
\bibliography{../literature,../references-mendeley}
%%\bibliographystyle{rsc} %the RSC's .bst file
%%\theendnotes
%}
%\appendix
%\section{Theorems}
%\label{sec:theorems}
%\begin{thm}[Berry-Esseen]
%  Let $X_1, X_2, \cdots$ be independent random variables with $E(X_i)=0$, $E(X_i^2)=\sigma_i^2>0$, $E(|X_i|^3)=\rho_i < \infty$. Also, let 
%  \begin{align*}
%    S_n = \frac{X_1 + X_2 + \cdots + X_n}{\sqrt{\sigma_1^2 +\sigma_2^2 + \cdots +\sigma_n^2}}
%  \end{align*}
%  be the normalized $n$-th partial sum. Denote $F_n$ the cdf of $S_n$, and $\Phi$ the cdf of the standard normal distribution. For the sake of convenience, denote 
%  \begin{align}
%    \vec \sigma = (\sigma_1, \cdots,\sigma_n), \; \vec \rho = (\rho_1, \cdots ,\rho_n)\,.
%  \end{align}
%  Then there exists and absolute constant $C_0$ such that
%  \begin{align}
%    &\sup_{x\in \mathbb R} |F_n(x) - \Phi(x)| \leq C_0 \cdot \psi_0\,, \\
%    \text{where} \quad &\psi_0 = \psi_0(\vec \sigma, \vec \rho) = \left(\sum_{i=1}^n \sigma_i^2\right)^{-3/2} \cdot \sum_{i=1}^n \rho_i\,.
%  \end{align}
%\end{thm}
%\begin{figure}
%  \centering
%  \includegraphics{length-distribution-n2}
%  \caption{Comparison between mean field result \cref{eq:meanfield} and simulations for a single regular spring network with $N=1000$ and $n=2$.}
%\end{figure}
%\begin{figure}
%  \centering
%  \includegraphics{length-distribution-n8}
%  \caption{Comparison between mean field result \cref{eq:meanfield} and simulations for a single regular spring network with $N=1000$ and $n=8$.}
%\end{figure}
\end{document}
%
% ****** End of file apssamp.tex ******